\documentclass{article}
\usepackage{graphicx}
\usepackage{amsmath}

\input{tcilatex}

\begin{document}

\title{Nonequilibrium Dynamics of Optical Lattice-Loaded BEC Atoms: Beyond HFB
Approximation}
\author{$^{1,2}$Ana Maria Rey , $^{1}$B. L. Hu, $^3$Esteban Calzetta, $^{1}$Albert
Roura, $^{2}$Charles Clark \\
$^{1}${\small Department of Physics, University of Maryland, College Park,
MD 20742}\\
$^{2}${\small National Institute of Standards and Technology, Gaithersburg,
MD 20899}\\
$^3${\small Departamento de Fisica, Facultad de Ciencias Exactas y Naturales,%
}\\
{\small Universidad de Buenos Aires- Ciudad Universitaria, 1428 Buenos
Aires, Argentina}}
\date{\today}
\maketitle

\begin{abstract}
In this work a two-particle irreducible (2PI) closed-time-path
(CTP) effective action is used to describe the nonequilibrium
dynamics of a Bose Einstein condensate (BEC) selectively loaded
into every third site of a one-dimensional optical lattice. The
motivation of this work is the recent experimental realization of
this system at National Institute of Standards and Technology
(NIST) where the placement of atoms in an optical lattice is
controlled by using an intermediate superlattice. This patterned
loading method is a useful technique for the proposed
implementation of lattice-based AMO quantum computing. This system
also serves to illustrate many basic issues in nonequilibrium
quantum field theory pertaining to the dynamics of quantum
correlations and fluctuations which goes beyond the capability of
a mean field theory. By numerically evolving in time the initial
state configuration using the Bose- Hubbard Hamiltonian an exact
quantum solution is available for this system in the case of few
atoms and wells. One can also use it to test out the various
approximation methods constructed. Under the 2PI CTP scheme with
this initial configuration, three different approximations are
considered: a) the \ Hartree-Fock-Bogoliubov (HFB) approximation,
b) the next-to-leading order 1/$\mathcal{N}$ expansion of the 2PI
effective action up to second order in the interaction strength
and c) a second order perturbative expansion in the interaction
strength. We present detailed comparisons between these
approximations and determine their range of validity by
contrasting them with the exact many body solution for a moderate
number of atoms and wells. As a general feature we observe that
because the second order 2PI approximations include multi-particle
scattering in a systematic way, they are able to capture damping
effects exhibited in the exact solution that a mean field
collisionless approach fails to produce. While the second order
approximations show a clear improvement over the HFB approximation
our numerical result shows that they do not work so well at late
times, when interaction effects are significant.
\end{abstract}


\bigskip

\section{Description of the Problem}

Bose-Einstein condensate (BEC) loaded into an optical lattice has
provided an interesting arena for the study of quantum coherence
and fluctuation phenomena  in many body physics. Spectacular
progress in experimental studies have been able to achieve regimes
where  standard mean field techniques used to describe weakly
interacting atoms are not generally applicable. The description of
the evolution of condensates far from equilibrium \ has also
gained considerable importance in matter-wave physics, motivated
by recent experimental realization of colliding and collapsing
condensates. In this paper we want to investigate the dynamics of
a Bose-Einstein condensate at zero-temperature (T=0) loaded every
three sites in a one dimensional optical lattice, recently
realized by the NIST group \cite{NISTLattice}.

In that experiment a combination of two independently controlled lattices
was used to prepare the condensate into every third site of a single
lattice. Briefly the procedure consists of loading a condensate into the
ground band of a lattice with periodicity $3a$, so that the condensate is
well localized in the potential minima of this lattice. A second lattice of
periodicity $a$ which is parallel to the first lattice is then ramped up so
that the superimposed light potentials form a super-lattice of period $3a$.
If both lattices are in phase, then the addition of the new lattice will not
shift the locations of the potential minima from those of the first lattice
alone and the condensate will remain localized at these positions. Finally,
by removing the first period$\ 3a$ lattice on a time scale long compared to
band excitations, but short compared to the characteristic time of dynamics
within the lattice, the condensate will be left in every third site of the
period $a$ lattice.

A condensate loaded in this way is not an eigenstate of the final period $a$
lattice, and the condensate will continue to evolve in the final system.
Outside the strongly correlated regime, a mean field approach was expected
to give a good description of the condensate dynamics \cite{Rey}. However,
by comparison with numerical simulations of the exact solution of the
many-body Hamiltonian it was found that even in the case when the kinetic
energy is comparable with the on-site interactions, inter-atomic collisions
play a crucial role in determining the quantum dynamics of the system and
therefore a mean field collisionless approach is only accurate for short
times.

Thus, in order to model the correct quantum dynamics of the system it is
necessary to include properly scattering processes among particles. This
task, however, is not easy for this particular system because contrary to
the three dimensional dilute gas case, where many body effects introduce
only a small change to the two-particle scattering properties in vacuum, the
presence of the lattice and the low dimensionality of the system make the
problem much less straightforward.

To date most theoretical descriptions of nonequilibrium dynamics
are based on the time dependent Gross-Pitaevskii equation coupled
with extended kinetic theories that describe excitations near
thermal equilibrium situations. These approaches usually rely on
the contact-potential approximation \ of the binary interactions \
which restricts two-body collisions to low energy. However, system
dynamics of most interest falls \ outside of \ this restricted
domain; its description thus requiring new methods. Here to
describe the far-from-equilibrium dynamics we adopt \ a CTP \
(Closed Time Path) \cite{ctp} functional-integral formalism
together with a 2PI (two-particle irreducible) \cite{2pi}
effective action approach to derive the equations \ of motion.
This method has been generalized for and applied to the
establishment of a quantum kinetic field theory \cite
{CH88,mea,CHR} with applications to problems in gravitation and
cosmology \cite{CH87,Ramsey}, particles and fields
\cite{CH89,qgp}, BEC \cite {Stoof2,Boyan} and condensed matter
systems \cite{RammerBook} as well as addressing the issues of
thermalization and quantum phase transitions \cite
{Cooper,Berges,Rivers}.

In Section 2 we summarize the mean field results obtained in
previous studies \cite{Rey}. In Section 3 \ we introduce the 2PI
generating functional to construct the 2PI effective action
$\Gamma_2$ and Green's functions. In Section 4 we perform
perturbative expansions on $\Gamma_2$ and define the various
approximation schemes. In Section 5 we introduce the CTP
formalism. \ We then derive the equation of motion and discuss the
results under each approximation scheme, starting with HFB in
Section 6, followed by second order expansions in Section 7, which
includes the 1/$\mathcal{N}$ expansion and the full second order
expansion. The numerical \ implementation is discussed in Section
8. In Section 9 we present our results and determine the range of
validity of the approximations by comparisons with the exact
(numerical) solution. We end with the conclusions.

\section{Mean Field Theory}

The dynamics of an ultracold \ bosonic gas in an optical lattice can be
described by a Bose-Hubbard model where the system parameters are controlled
by laser light. For a one dimensional lattice the starting Hamiltonian is:

\begin{equation}
\hat{H}=-J\sum_{i}(\hat{\Phi}_{i\;}^{\dagger }\hat{\Phi}_{i+1\;}+\hat{%
\Phi}_{i+1\;}^{\dagger }\hat{\Phi}_{i\;})+\sum_{i}\epsilon _{i}\hat{%
\Phi}_{i\;}^{\dagger }\hat{\Phi}_{i\;}+\frac{1}{2}U\sum_{i}\hat{\Phi%
}_{i\;}^{\dagger }\hat{\Phi}_{i\;}^{\dagger }\hat{\Phi}_{i\;}\hat{%
\Phi}_{i\;},  \label{BHH}
\end{equation}
where $\hat{\Phi}_{i}$ and $\hat{\Phi}_{i}^{\dagger }$ \ are the
annihilation and creation operators at the site $i$ \ which obey
the canonical commutation relations for bosons. Here, the
parameter $U$ denotes the strength of the on-site repulsion of two
atoms on the site $i$; the parameter $\varepsilon _{i}$ denotes
the energy offset of each lattice site due to an additional slow
varying external potential such as a magnetic trap and $J$ denotes
the hopping rate between adjacent sites. Because the
next-to-nearest neighbor amplitudes are \ typically two orders of
magnitude smaller, tunneling to them \ can be neglected. The Bose
-Hubard Hamiltonian should be an appropiate model when the loading
process produces atoms in the lowest vibrational state of each
well, with a chemical potential \ smaller than the distance \ of
the first vibrationally excited state. This is the case of the
experiment we want to study\cite{NISTLattice}.

Here we consider only a one-dimensional homogeneous lattice with periodic
boundary conditions. The transverse degrees of freedom and the harmonic
confinement in the direction of the lattice are relegated to a future \ work.

In the simplest version of the mean-field theory, we assume that all atoms
share the same condensate wave function and therefore it is possible to
replace the operator on the lattice site $i$ by a c-number $\phi _{i}(t)$.
Under this approximation $\phi _{i}(t)$ corresponds to the time dependent
mean value amplitude of the $i^{th\text{ }}$ well. One might expect that
this zero temperature mean field approach could give a good description of
the dynamics of the atoms in the limit that quantum fluctuations can be
completely neglected, that is, the regime when the ratio U to J is
sufficiently small and the tunneling overwhelms the repulsion.

By making the mean field ansatz in the Bose- Hubbard Hamiltonian
it is possible to show that the $\phi _{i}(t)$ amplitudes satisfy
the Discrete Non Linear Schrodinger Equation (DNLSE), which in the
case of zero external potential has the form:

\begin{equation}
i \frac{\partial \phi _{i}}{\partial \tau}=-(\phi _{i-1}+\phi _{i+1})+\frac{%
UN}{J}|\phi _{i}|^{2}\phi _{i},  \label{tres}
\end{equation}
where we have assumed that the $\phi _{i}(\tau)$ are normalized to one, that
$N$ is the total number of atoms and defined a new time scaled $\tau\equiv%
\frac{t J}{\hbar}$.

We treat a model case in which the initial occupancies of each third site
are the same, and in which the condensate initially has a uniform phase.
Thus at $t=0$, the amplitudes $\phi _{i}(\tau)$ are given by $\phi _{3i}(0)=%
\sqrt{I/3},$ $\phi _{3i+1}(0)=\phi _{3i+2}(0)=0$, where $I$ \ is the total
number of lattice sites. For an infinite lattice, or one with periodic
boundary conditions, the amplitudes for all initially occupied sites $\phi
_{3i}$ evolve identically in time, and the amplitudes for the initially
unoccupied sites satisfy $\phi _{3i+1}(\tau)=\phi _{3i+2}(\tau)$ for all $%
\tau$. This allows us to reduce the full set of equations (\ref{tres}) to a
set of two coupled equations for $\phi _{0}(\tau)$ and $\phi _{1}(\tau)$.

The solutions $|\phi _{0}(\tau)|$ and $|\phi _{1}(\tau)|$ are oscillatory
functions whose amplitudes and common period, $T(\gamma )$, are determined
by the parameter $\gamma \equiv NUI/(3J)$. It is useful to qualitatively
divide the dynamical behavior into two regimes \cite{Rey}:

\paragraph{The tunneling dominated regime $(\protect\gamma <1)$:}

In this regime we find that the oscillation period is essentially constant,
the role of interactions is relatively small, and the equations of motion
are equivalent to those of a two-state Rabi problem. This system will
undergo Rabi oscillations whereby atoms periodically tunnel from the
initially occupied site into the two neighboring sites. For $\gamma =0$ the
period of oscillation is $\frac{2\pi }{3}$.

\paragraph{Interaction dominated regime:}

The effect of interactions on the mean-field dynamics is to cause the
energies of the initially occupied sites to shift relative to those of the
unoccupied sites. As $\gamma $ increases the tunneling between sites occurs
at a higher frequency, but with reduced amplitude. The population of the
initially occupied sites becomes effectively self-trapped by the purely
repulsive pair interaction.

To check the validity of the mean field approximation, we made comparisons
with the exact many body solution for 6 atoms and three wells. We use a
modest number of atoms and lattice sites for the comparisons \ due to the
fact that the Hilbert space needed for the calculations increases rapidly
with the number of atoms and wells. The exact solution was obtained by
evolving an initial state $(e^{-N/2}$ $e^{\sqrt{N}\hat{\Phi _{o}^{\dagger }}%
}\left| 0\right\rangle )\otimes $ $\left| 0\right\rangle $
$\otimes \left| 0\right\rangle $ with the Bose-Hubbard
Hamiltonian. The initial state represents just a coherent state
with an average of $N$ atoms in the initial populated well. (See
Sec. 8.1 ).

In Fig. 1 we plot the average population per well $\ \left\langle \hat{%
\Phi}_{i\;}^{\dagger }(t)\hat{\Phi}_{i\;}(t)\right\rangle $ and
the
condensate population per well $\left| \left\langle \hat{\Phi}%
_{i\;}(t)\right\rangle \right| ^{2}$ and compare them with the mean field
predictions, i.e. \ $|\phi _{i}(t)|^{2}$, for three different values of $%
\gamma$. The salient features observed in these comparisons are:

\paragraph{Weakly interacting regime}

($\gamma =0.2$): The DNLSE gives a good description of the early
time dynamics. The total population per well predicted by the mean
field solution agrees with the exact solution. This is because the
total condensate population remains big for the time in
consideration.

\paragraph{Intermediate regime}

($\gamma =2$): Quantum fluctuations lead to a non-trivial
modulation of the classical oscillations. In this regime the ratio
between interaction and kinetic energy is small enough to allow
the atoms to tunnel but not too small to make interaction effects
negligible, the mean field results are accurate only for a short
time. The exact solution exhibits for this regime damped
oscillations of the atomic population per well that drives the
system towards equilibrium. There is also a very rapid reduction
of the condensate population. These characteristic damping effects
are \ caused by the different phase evolution \ of \ the quantum
states that span the configuration space and collisions among
atoms. Quantum scattering effects are crucial, even for rather
early times.

\paragraph{Strongly correlated regime}

($\gamma =12$): the system exhibits \ macroscopic quantum
self-trapping of the population. Qualitatively, both \ the mean
field and the exact solutions, agree in the sense that atoms are
in both self-trapped in the initially populated wells due to
interactions. However, the fast decrease of the condensate
population and its subsequent revivals shown in the exact
solutions give us an idea of the importance of correlation effects
beyond mean field in this interaction dominated regime. Due to the
fact that the evolution of the initially unoccupied wells is
almost frozen by \ interactions, \ the collapse and revival of the
condensate population can be mainly described by the time
evolution of the n-particle number states
that expand the coherent state of the initially populated well. \footnote{%
The partial revival of the condensate population can be understood
as \ the interference of states approximated by a quadratic
spectrum: $E_{n}\approx U/2n(n-1)$. At first, the different phase
evolution of the \ n-particle states leads to a collapse of the
condensate population. However at integer values of $\ t_{rev}=\
(U/h)^{-1}$ \ the phase factors add to an integer value of $2\pi
$, leading to a \ revival of the initial state \cite{Bloch}. The
revival is not complete because $J$ is not exactly zero. The
collapse time $t_{c}$ depends on the variance of the initial
atomic distribution $\ t_{c}\approx \ (t_{rev}/\sigma )$ (See
Ref.~ \cite{Smerzi}-\cite{Walls}).
For the initial Poissonian distribution \ $\ t_{c}\approx (U\sqrt{N}%
/h)^{-1}. $}

Because our main interest is the tunneling dynamics we will focus on the
intermediate regime, where the ratio U/J is small but interaction effects
are not negligible. In this regime a perturbative expansion around U/J still
makes sense.

\begin{figure}[tbh]
\begin{center}
\leavevmode {\includegraphics[width=3.5 in]{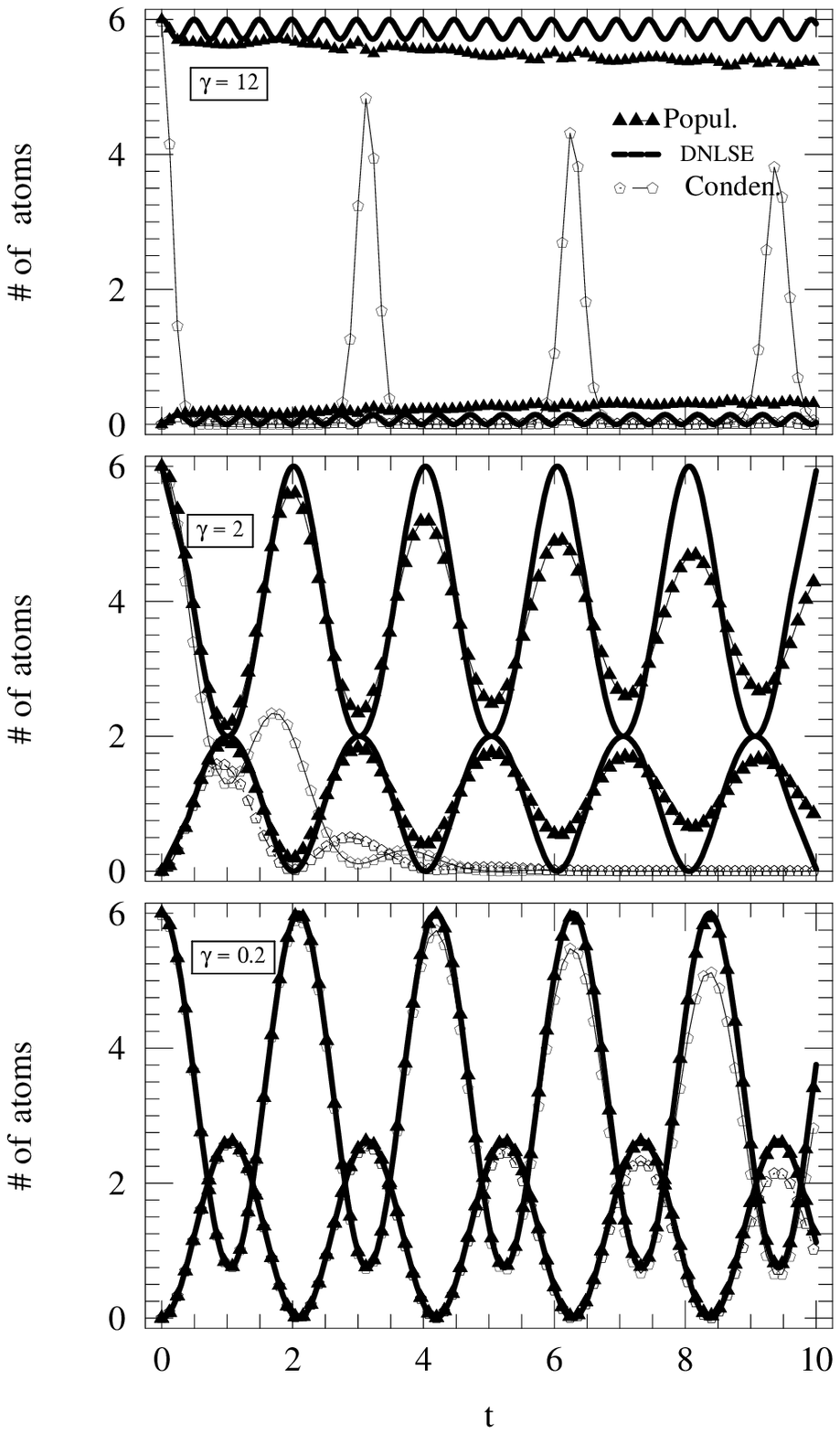}} \
\end{center}
\caption{Comparisons between the exact and the DNLSE solutions for six atoms
and three wells. The time is given in units of $\hbar/J$. Top panel,
strongly correlated regime ($\protect\gamma=12$); Middle panel, intermediate
regime ($\protect\gamma=2$) and bottom panel weak interacting regime($%
\protect\gamma=0.2$). The solid line is the DNLSE prediction for the
population per well: $|\protect\phi_{0}(t)|^{2}$ and $|\protect\phi%
_{1,2}(t)|^{2}$ (see Eq. \ref{tres}), the triangles are used to represent
the exact solution for the population per well calculate using the Bose
Hubbard Hamiltonian (Eq. \ref{BHH}): $\langle\hat{\protect\Phi}%
_{0}^{\dagger }\hat{\protect\Phi}_{0}\rangle$, $\langle\hat{\protect%
\Phi}_{1,2}^{\dagger }\hat{\protect\Phi}_{1,2}\rangle$. The
pentagons
show the condensate population per well calculated from the exact solution: $%
|\langle\hat{\protect\Phi}_{0}\rangle|^{2}$ and $|\langle\hat{\protect%
\Phi}_{1,2}\rangle|^{2}$. Due to the symmetry of the initial
periodic conditions the curves for the $i=1$ and $2$ wells are the
same in all depicted curves .} \label{Fig1}
\end{figure}

\section{2PI Effective Action $\Gamma (\protect\phi ,G)$}

\begin{figure}[tbh]
\begin{center}
\leavevmode {\includegraphics[width=4.2 in]{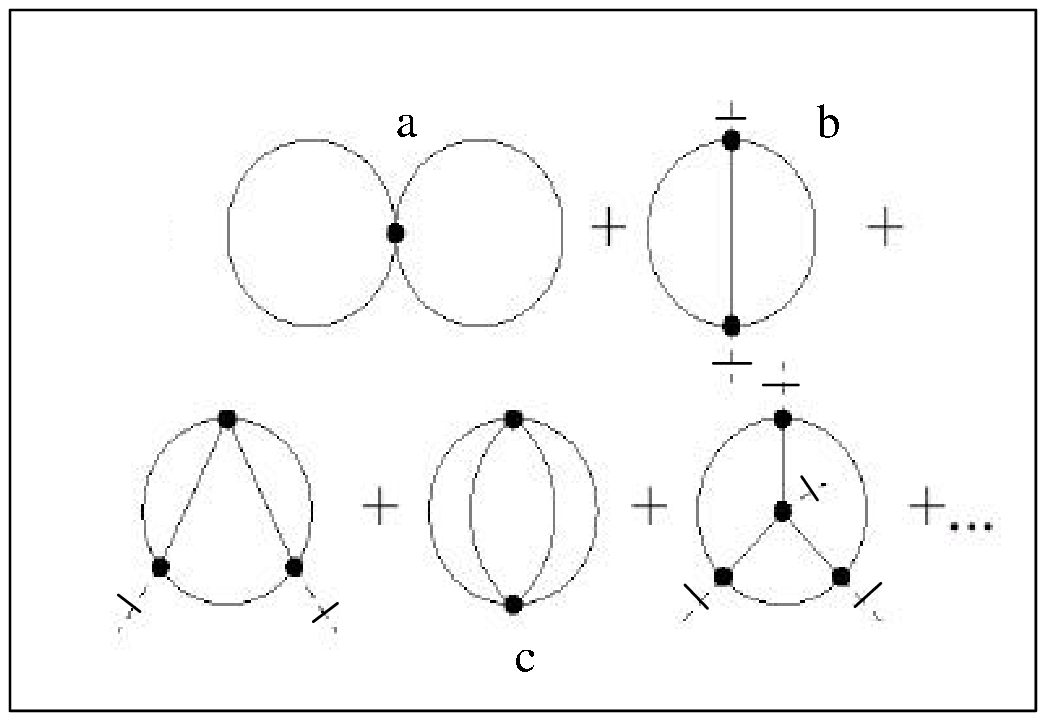}}
\end{center}
\caption{Two-loop (upper row) and three-loop diagrams (lower row)
contributing to the effective action. Explicitly, the diagram a is what we
call the \emph{double-bubble} , b the \emph{setting-sun} and c the \emph{%
basket-ball}. }
\label{Fig2}
\end{figure}

The first requirement for the study of nonequilibrium processes is a general
initial-value formulation depicting the dynamics of interacting quantum
fields. The CTP or Schwinger-Keldysh effective action formalism \cite{ctp}
serves this purpose. The second requirement is to describe the evolution of
the correlation functions and the mean field on an equal footing. The two
particle \ irreducible (2PI) formalism \cite{2pi} where the correlation \
functions appear also as independent variables\ serves this purpose. By
requiring the generalized (master) CTP effective action \cite{mea} to be
stationary with respect to variations \ of the correlation functions \ an
infinite set of coupled (Schwinger-Dyson) equations for the correlation
functions is obtained which is a quantum analog of the BBGKY hierarchy. The
2PI effective action produces two such functions in this hierarchy. In this
section we shall focus on the 2PI formalism, and then upgrade it to the CTP
version in the next section.

An adequate description of the dynamics of atoms loaded into an
optical lattice is provided by the Bose-Hubbard Hamiltonian
(\ref{BHH}), whose classical action is given in terms of the
complex fields $\Phi _{i}^{{}}$ and $\Phi _{i}^{\ast }$ by

\begin{equation}
S[\Phi _{i}^{\ast },\Phi _{i}]=\int dt\sum_{i}i\hbar \Phi
_{i}^{\ast }(t)\partial _{t}\Phi _{i}(t)+J\left( \Phi _{i}^{\ast
}(t)\Phi _{i+1}(t)+\Phi _{i}(t)\Phi _{i+1}^{\ast }(t)\right) -\frac{%
U}{2}\Phi _{i}^{\ast }(t)\Phi _{i}^{\ast }(t)\Phi _{i}(t)\Phi
_{i}(t),
\end{equation}
where, as before, $i$ denotes the lattice position, $\ J$ is the hopping
rate and $U$ is the interaction strength. We limit the analysis to the case
when no external potential is present and include only nearest neighbor
hopping.

To compactify our notation we introduce $\Phi _{i}^{a} (a=1,2)$
defined by
\begin{equation}
\Phi _{i} =\Phi _{i}^{1}, \;\; \Phi _{i}^{\ast } =\Phi _{i}^{2}.
\end{equation}
In terms of these fields the classical action takes the form

\begin{equation}
S[\Phi ]=\int dt\sum_{i}\frac{1}{2}h_{ab}\Phi _{i}^{a}(t)\hbar
\partial _{t}\Phi _{i}^{b}(t)+J\sigma _{ab}\Phi _{i+1}^{a}(t)\Phi
_{i}^{b}(t)-\frac{U}{4\mathcal{N}}(\sigma _{ab}\Phi
_{i}^{a}(t)\Phi _{i}^{b}(t))^{2},
\end{equation}
where $\mathcal{N}$ is the number of fields,which is two in this case, and summation over repeated field indices $%
a,b=(1,2)$ is implied. $h_{ab}$ and $\sigma _{ab}$ are matrices defined as

\begin{equation}
h_{ab}=i\left(
\begin{array}{cc}
0 & -1 \\
1 & 0
\end{array}
\right) \qquad \sigma _{ab}=\left(
\begin{array}{cc}
0 & 1 \\
1 & 0
\end{array}
\right)
\end{equation}

After second quantization the fields $\Phi _{i}^{a}$ are promoted
to operators. We denote the expectation value of the field
operator or mean field \ by $\phi _{i}^{a}(t)$ and the expectation
value\ of the composite field \ by $G_{ij}^{ab}(t,t^{\prime })$.
Physically, $|\phi _{i}^{a}(t)|^{2}$ is the condensate population
and the composite fields determine the fluctuactions around the
mean field.

\begin{eqnarray}
\phi _{i}^{a}(t) &=&\left\langle \Phi _{i}^{a}(t)\right\rangle, \\
\hbar G_{ij}^{ab}(t,t^{\prime }) &=&\left\langle T_{C}\Phi
_{i}^{a}(t)\Phi _{i}^{b}(t^{\prime })\right\rangle -\left\langle
\Phi _{i}^{a}(t)\right\rangle \left\langle \Phi _{i}^{b}(t^{\prime
})\right\rangle.
\end{eqnarray}
The brackets denote taking the expectation value respect to the density matrix and $%
T_{C}$ denotes time ordering along a contour C in the complex plane.

\bigskip

All correlation functions of the quantum theory can be obtained from the
effective action $\Gamma \lbrack \phi ,G]$, the two particle irreducible
generating functional \ for Green's functions parametrized by $\phi
_{i}^{a}(t)$ and the composite field $G_{ij}^{ab}(t,t^{\prime })$. To get an
expression for the effective \ action we first define \ the functional $Z[%
\mathbf{J,K}]$ \cite{2pi} as

\begin{eqnarray}
Z[\mathbf{J,K}] &=&e^{i/\hbar W[\mathbf{J,K}]}  \label{zeq} \\
&=&\prod_{a}\int D\Phi ^{a}\exp \left\{ \frac{i}{\hbar }\left(
S[\Phi ]+\int dt\sum_{i}\mathbf{J}_{ia}(t)\Phi
_{i}^{a}(t)+\frac{1}{2}\int
dtdt^{\prime }\sum_{ij}\Phi _{i}^{a}(t)\Phi _{j}^{b}(t^{\prime })%
\mathbf{K}_{ijab}(t,t^{\prime })\right) \right\} ,  \notag
\end{eqnarray}
where we have introduced the following index lowering convention
\begin{equation}
X_{a}=\sigma _{ab}X^{b}.
\end{equation}
The functional integral (\ref{zeq}) is a sum over classical histories of the
field $\Phi _{i}^{a}$ \ in the presence of \ the local source \ $\mathbf{J%
}_{ia}$ and the non local source $\mathbf{K}_{ijab}.$ The coherent
state measure is included in $D\Phi $. The addition of \ the
two-particle source term is what characterizes the 2PI formalism.

We define $\Gamma \lbrack \phi ,G]$ as the double Legendre transform of $W[%
\mathbf{J,K}]$ such that

\begin{eqnarray}
\frac{\delta W[\mathbf{J,K}]}{\delta \mathbf{J}_{ia}(t)} &=&\phi _{i}^{a}(t),
\\
\frac{\delta W[\mathbf{J,K}]}{\delta \mathbf{K}_{ij ab}(t,t^{\prime })} &=&%
\frac{1}{2}[\phi _{i}^{a}(t)\phi _{i}^{b}(t^{\prime })+\hbar
G_{ij}^{ab}(t,t^{\prime })].
\end{eqnarray}
Expressing $\mathbf{J}$ and $\mathbf{K}$ in terms of $\phi $ and $G$ \ yields

\begin{eqnarray}
\Gamma \lbrack \phi ,G] &=&W[\textbf{J},\textbf{K}]-\int
dt\sum_{i}\mathbf{J}_{ia}(t)\phi _{i}^{a}(t)-\frac{1}{2}\int
dtdt^{\prime }\sum_{ij}\phi _{i}^{a}(t)\phi
_{j}^{b}(t^{\prime })\mathbf{K}_{ijab}(t,t^{\prime }) \\
&&-\frac{\hbar }{2}\int dtdt^{\prime }\sum_{ij}G_{ij}^{ab}(t,t^{\prime })%
\mathbf{K}_{ijab}(t,t^{\prime }). \label{gamm} \notag
\end{eqnarray}
{}From this equation the following identity can be derived:

\begin{eqnarray}
\frac{\delta \Gamma \lbrack \phi ,G]}{\delta \phi _{i}^{a}(t)} &=&-\left(
\mathbf{J}_{ia}(t)+\int dt^{\prime }\sum_{j}(\mathbf{K}_{ijad}(t,t^{\prime
}))\phi _{j}^{d}(t^{\prime })\right) ,  \label{eaeq1} \\
\frac{\delta \Gamma \lbrack \phi ,G]}{\delta G_{ij}^{ab}(t,t^{\prime })} &=&-%
\frac{\hbar }{2}\mathbf{K}_{ijab}(t,t^{\prime }).  \label{eaeq2}
\end{eqnarray}

In order to get an expression for $\Gamma \lbrack \phi ,G]$ notice
that by using Eq.(\ref{zeq}) for $W[\textbf{J}, \textbf{K}]$ and
placing it in Eq.(\ref{gamm}) for $\Gamma \lbrack \phi ,G]$, it
can be written as

\begin{eqnarray}
\exp \left( \frac{i}{\hbar }\Gamma \lbrack \phi ,G]\right)  &=&\prod_{a}\int
D\Phi ^{a}\exp \left\{ \frac{i}{\hbar }\left( S[\Phi ]+\int dt_{i}\;%
\mathbf{J}_{ia}(t)\left[ \Phi _{i}^{a}(t)-\phi _{i}^{a}(t)\right]
\right.
\right.  \\
&&\left. \left. +\frac{1}{2}\int dt_{i}dt_{j}^{\prime }\left( \Phi
_{i}^{a}(t)\mathbf{K}_{ijab}^{{}}(t,t^{\prime })\Phi
_{j}^{b}(t^{\prime })-\phi
_{i}^{a}(t)\mathbf{K}_{ijab}^{{}}(t,t^{\prime })\phi
_{j}^{b}(t^{\prime })\right) -\frac{\hbar }{2}TrG\mathbf{K}\right)
\right\}
\notag \\
&=&\prod_{a}\int D\Phi ^{a}\exp \left\{ \frac{i}{\hbar }\left(
S[\Phi
]-\int dt_{i}\;\frac{\delta \Gamma \lbrack \phi ,G]}{\delta \phi _{i}^{a}(t)}%
\left[ \Phi _{i}^{a}(t)-\phi _{i}^{a}(t)\right] \right. \right.
\notag
\\
&&\left. \left. -\frac{1}{\hbar }\int dt_{i}dt_{j}^{\prime
}\;\left[ \Phi
_{i}^{a}(t)-\phi _{i}^{a}(t)\right] \frac{\delta \Gamma \lbrack \phi ,G]}{%
\delta G_{ij}^{ab}(t,t^{\prime })}\left[ \Phi _{i}^{b}(t^{\prime
})-\phi
_{i}^{b}(t^{\prime })\right] +TrG\frac{\delta \Gamma \lbrack \phi ,G]}{%
\delta G}\right) \right\} ,  \notag
\end{eqnarray}
where Tr means taking the trace. For simplicity we have denoted
$\int dt\sum_{i}$ by $\int dt_{i}$. Defining the fluctuation
field, $\varphi_{i}^{a}=\Phi_{i}^{a}-\phi_{i}^{a}$, we have

\begin{eqnarray}
\Gamma \lbrack \phi ,G]-TrG\frac{\delta \Gamma \lbrack \phi ,G]}{\delta G}
&=&-i\hbar \ln \prod_{a}\int D\varphi ^{a}\exp \left( \frac{i}{\hbar }S[\phi
,G;\varphi ]\right)  \label{eact} \\
S[\phi ,G;\varphi ] &=&S[ \phi +\varphi]-\int dt_{i}\;\frac{\delta
\Gamma \lbrack \phi ,G]}{\delta \phi _{i}^{a}(t)}\varphi
_{i}^{a}(t)-\frac{1}{\hbar }\int dt_{i}dt_{j}^{\prime }\;\varphi
_{i}^{a}(t)\frac{\delta \Gamma \lbrack \phi ,G]}{\delta
G_{ij}^{ab}(t,t^{\prime })}\varphi _{i}^{b}(t^{\prime }).
\label{shift}
\end{eqnarray}
By introducing the classical inverse propagator $iD^{-1}(\phi )$ given by

\begin{eqnarray}
iD_{ijab}(t,t^{\prime })\;^{-1} &=&\frac{\delta S[\phi ]}{\delta \phi
_{i}^{a}(t)\delta \phi _{j}^{b}(t^{\prime })} \\
&=&\left( \delta _{ij}h_{ab}\partial _{t}+J(\delta _{i+1j}+\delta
_{i-1j})\sigma _{ab}\right) \delta (t-t^{\prime })  \notag \\
&&-\frac{U}{\mathcal{N}}\left( 2\phi _{ia}(t)\phi _{ib}(t)+\sigma _{ab}\phi
_{i}^{c}(t)\phi _{ic}(t)\right) \delta _{ij}\delta (t-t^{\prime }),  \notag
\end{eqnarray}
the solution of the functional integro-differential equation (\ref{eact})
can be expressed as

\begin{equation}
\Gamma \lbrack \phi ,G]=S[\phi ]+\frac{i}{2}Tr\ln G^{-1}+\frac{i}{2}%
TrD^{-1}(\phi )G+\Gamma _{2}[\phi ,G]+const.  \label{2pi}
\end{equation}
The quantity $\Gamma _{2}[\phi ,G]$ is conveniently \ described in
terms \ of the diagrams generated by the \ interaction terms in
$S[ \phi +\varphi]$ which are \ of cubic and higher orders in
$\varphi $.

\begin{equation}
S_{int}[ \phi +\varphi]=-\frac{U}{4\mathcal{N}}\int
dt_{i}\;(\varphi _{ib}(t)\varphi
_{i}^{b}(t))^{2}-\frac{U}{\mathcal{N}}\int dt_{i}\;\varphi
_{i}^{a}(t)\phi _{ia}(t)\varphi _{i}^{b}(t)\varphi _{ib}(t).
\end{equation}
It consists of all \ two-particle irreducible vacuum graphs \ (the
diagrams representing these interactions do not become
disconnected by cutting two propagator lines) in the theory with
propagators set equal to $G$ and vertices determined by \ the \
interaction terms in $S[ \phi +\varphi]$ .

Since physical processes correspond to \ vanishing sources $\mathbf{J}$ and $%
\mathbf{K}$, \ the dynamical equations of motion for the mean field and the
propagators are found \ by using the expression (\ref{2pi}) in equations (%
\ref{eaeq1})\ and (\ref{eaeq2}) , and setting the right hand side equal to
zero. This procedure leads to the following equations:

\begin{eqnarray}
h_{ab}\hbar \partial _{t}\phi _{i}^{b}(t) &=&-J(\phi _{i+1a}(t)+\phi
_{i-1a}(t))+\frac{U}{\mathcal{N}}(\phi _{id}(t)\phi
_{i}^{d}(t)+G_{ii\;c}^{\;c}(t,t))\phi _{ia}(t)+  \label{cond} \\
&&\frac{U}{\mathcal{N}}((G_{iiad}(t,t)+G_{iida}(t,t))\phi _{i}^{d}(t)-\frac{%
\delta \Gamma _{2}[\phi ,G]}{\delta \phi _{i}^{a}(t)},  \notag
\end{eqnarray}

\noindent and

\begin{eqnarray}
G_{ijab}^{-1}(t,t^{\prime }) &=&D_{ijab}(t,t^{\prime })^{-1}-\Sigma
_{ijab}(t,t^{\prime }),  \label{inv} \\
\Sigma _{ijab}(t,t^{\prime }) &\equiv &2i\frac{\delta \Gamma _{2}[\phi ,G]}{%
\delta G_{ij}^{ab}(t,t^{\prime })} .  \label{pder}
\end{eqnarray}
Equation\ (\ref{inv}) can \ be rewritten \ as a partial differential
equation \ suitable for initial value problems \ by convolution with $G$.
This differential equation reads explicitly

\begin{eqnarray}
h_{c}^{a}\hbar \partial _{t}G_{ij\;}^{\;cb}(t,t^{\prime })
&=&-J(G_{i+1j}^{ab}(t,t^{\prime })+G_{i-1j}^{ab}(t,t^{\prime }))+\frac{U}{%
\mathcal{N}}(\phi _{id}(t)\phi _{i}^{d}(t))G_{ij}^{ab}(t,t^{\prime })+
\label{fluc} \\
&&\frac{2U}{\mathcal{N}}\phi _{i}^{a}(t)G_{ij}^{cb}(t,t^{\prime })\phi
_{ic}(t)+i\int dt_{k}^{\prime \prime }\Sigma _{ikc}^{\;a}(t,t^{\prime \prime
})G_{kj}^{cb}(t^{\prime \prime },t^{\prime })+i\delta ^{ab}\delta
_{ij}\delta (t-t^{\prime }).  \notag
\end{eqnarray}
The evolution of $\phi ^{a}$ and $G^{ab}$ is determined by Eqs. (\ref{cond})
and (\ref{fluc}) once $\Gamma _{2}[\phi ,G]$ is specified.

\section{Perturbative Expansion of $\Gamma _{2}(\protect\phi ,G)$ and
Approximation Schemes}

\bigskip The diagrammatic expansion of \ $\Gamma _{2}$ is illustrated in
Fig. 2, where two and three-loop vacuum diagrams are shown. The dots where
four lines meet represent interaction vertices. \ The expression \
corresponding to each vacuum diagram should be multiplied by a factor $%
(-i)^{l}(i)^{s-2}$ where $l$ is the number of solid lines and \ s the number
of loops the diagram contains.

The action \ $\Gamma $ including the full diagrammatic series for $\Gamma
_{2}$ gives the full dynamics. It is of course not feasible to obtain an
exact expression for $\Gamma _{2}$ in a closed form. Various approximations
for the full 2PI effective action can be obtained by truncating the
diagrammatic expansion \ for $\Gamma _{2}$. Which approximation is most
appropriate depends on the physical problem under consideration:

\subsection{The standard approaches}

\begin{enumerate}
\item  \ Bogoliubov (One-loop) Approximation:

The simplest approximation consists of discarding $\Gamma _{2}$ altogether.
This yields the so called Bogoliubov or \ one-loop approximation whose
limitations have been extensively documented in the literature (\cite
{Morgan,BurnettComparisons}).

\item  Time-dependent Hartree-Fock-Bogoliubov (HFB) Approximation:

A truncation of $\Gamma _{2}$ retaining only the first order diagram in U,
i.e., keeping only the \emph{double- bubble} diagram, Fig. $2a$, yields
equations of motion of $\phi $ and $G$ which correspond to the time
dependent Hartree-Fock-Bogoliubov (HFB) approximation. This approximation
violates Goldstone's theorem, but conserves energy and particle number \cite
{Martin,Griffin,Giorgini}). The HFB equations can also be obtained by using
cumulant expansions up to the second order \cite{Burnett} in which all
cumulants containing three or four field operators \ are neglected. The HFB
approximation neglects multiple scattering. It can be interpreted as an
expansion in terms of $Ut/J,$ (where $t$ is the time of evolution) and is
good for the description of \ short time dynamics or weak interaction
strengths. It will be described in Sec. 6.
\end{enumerate}

\subsection{Higher order expansions}

We make a few remarks on the general properties of higher order expansions
and then specialize to two approximations.

\paragraph{2PI and Ladder Diagrams}

Since the work of Beliaev's \cite{Beliaev} and Popov \cite{Popov}
it is well known in the literature (see for example
Ref.~\cite{Stoof2,Stoof}) that including higher order terms in a
diagrammatic expansion corresponds to renormalizing the bare \
interaction potential to the four-point vertex, thus accounting
for the repeated scattering of the bosons. In Ref.~\cite {Rammer}
the authors have shown explicitly for a homogeneous Bose gas that
taking into account the two-loop contribution to the 2PI effective
action leads to diagrams topologically identical to those found by
Beliaev \ but with the exact propagator instead of the one-loop
propagator. \ In the dilute gas limit, where the inter-particle
distance is large compared with the s-wave scattering length, the
ladder diagrams give the largest contribution to the four-point
vertex. \ Every rung in a ladder contributes to a factor
proportional to $Um$
(In the presence of the lattice m should be replaced by the effective mass $%
m^{\ast }$ ( $m^{\ast }\sim \hbar ^{2}/(Ja^{2})$) with $a$ the
lattice spacing). The ladder resummation results in an effective
potential \ which is called the T- matrix. To lowest order in the
diluteness parameter , the T-matrix in 3-D systems can be
approximated by a constant proportional to the scattering length
(pseudopotential approximation). However this approximation is
only valid \ in the weak interaction limit and neglects all
momentum dependence which appear in the problem as higher order
terms. In that sense the 2PI effective action approach allows us
to go beyond the weakly interacting limit in a systematic way \
and to treat collisions more accurately.

\paragraph{Nonlocal Dissipation and Non-Markovian Dynamics }

Higher order terms  lead to nonlocal equations and dissipation. The presence
of nonlocal terms in the equations of motion \ is a consequence of the fact
that the 2PI effective action really corresponds \ to a further
approximation \ of the master effective equation \cite{mea}. The 2PI
effective action \ is obtained by the \emph{slaving} of the three point
function \ $C_{3}$ to the mean field and $G$ with a particular choice of
boundary conditions. See Ref. \cite{mea} for further details.

Non-Markovian dynamics is a generic feature of the nPI formalism
which yields integro-differential equations of motion. This makes
numerical solution difficult, but is a necessary price to pay for
a fuller account of the quantum dynamics. Many well acknowledged
approaches to the quantum kinetics of such systems adopt either
explicitly or implicitly (or at the end what amounts to) a
Markovian approximation \cite{Holland}. It assumes that only the
current configuration of the system, but not its history,
determines its future evolution. Markovian approximations are made
if one assumes instantaneous interactions, or in the kinetic
theory context that the time scales  between the duration of
binary collisions $\tau _{o}$\ and the inverse collision rate
$\tau _{c}$ are well-separated. In the low kinetic energy, weak
interacting regime the time between collisions (or mean free path)
\ is long compared to the reaction time (or scattering length):\
$\tau _{c}>>\tau _{o}$. The long separation between collisions \
and the presence of intermediate weak fluctuations, \ allow for a
rapid decay of the temporal and spatial correlations created \
between collision partners, which one can use to justify the
Markovian approximation. However, in the problem at hand, owing to
the presence of the lattice which confines the atoms to the bottom
of the wells with enhanced interaction effects, the \ low
dimensionality of the system and the far-from-equilibrium initial
conditions, non-Markovian dynamics needs to be confronted
squarely. That is the rationale for our adoption of the CTP 2PI
scheme. Now, for the specifics:

\begin{enumerate}
\item  Second Order expansion:

A truncation retaining diagrams of second order in $U$ containing besides
the \emph{double-bubble}, also the \emph{setting-sun} and the \emph{%
basket-ball} Fig.2. By including the \emph{setting-sun} and the \emph{%
Basket-ball} in the approximations we are taking into account two particle
scattering processes \cite{Ramsey,CHR}. Second order terms lead to integro-
differential equations which depend on the time history of the system.

\item  Large-$\mathcal{N}$ approximation

The 1/$\mathcal{N}$ expansion is a controlled \ non-perturbative
approximation \ scheme which \ can be used to study non-equilibrium quantum
field \ dynamics in the regime \ of strong \ interactions\cite{Berges,Cooper}%
. In the large $\mathcal{N}$ approach the field is modeled by $\mathcal{N}$
fields and the quantum field generating functional is expanded \ in powers
of 1/$\mathcal{N}$. \ In this sense the method is a controlled expansion in
a small parameter but \ unlike perturbation theory in the coupling constant,
which corresponds to an expansion around the vacuum, the large $\mathcal{N}$
expansion corresponds to an expansion of the theory about a strong
quasiclassical field.\
\end{enumerate}

In this work we will derive the equations of motion and perform numerical
calculations up to the second order in the coupling constant $U$. This will
enable us to determine the range of validity of three types of
approximations described above, namely, a) the HFB, b) the full second order
and c) NLO large $\mathcal{N}$ expansion up to second order in $U$ (in the
figures we use the shorthand HFB, 2nd and 1/$\mathcal{N}$ respectively) by
comparison with the exact many body solution for a moderate number of atoms
and wells.

\section{CTP formalism}

In order to describe the non-equilibrium dynamics we will now specify the
contour of integration in Eqs. (\ref{cond}) and  (\ref{fluc}) to be the
Schwinger-Keldysh contour \cite{ctp} along the real-time axis or \emph{%
closed time path } (CTP) contour. \ The Schwinger-Keldysh
formalism is a powerful method for deriving real and causal
evolution equations for the expectation values \ of quantum
operators for nonequilibrium fields. The basic idea of the CTP
formalism relies on the fact that a diagonal matrix element of a
system \ at a given time, $t=0$ can be expressed as a product of
transition matrix elements from $t=0$ to $t^{\prime }$ and the
time-reverse (complex conjugate) matrix element from $t^{\prime }$
to $0$ by inserting a complete set of states \ into this matrix
element at the later time $t^{\prime }$. Since each term in the
product is a transition matrix element \ of the usual or time
reversed kind, the standard path integral representation for each
one can be introduced. However, to get the generating functional
we seek, we have to require that the forward time evolution takes
place in the presence of a source $J^{+}$ but the reversed time
evolution takes place \ in the presence of a different source
$J^{-}$, otherwise all the dependence on the source drops out.

The doubling of sources, the fields \ and integration contours \ suggest
introducing \ a 2 x 2 matrix notation. This notation has been discussed
extensively in the literature (see \cite{CH88,CHR}). However we are going \
to follow Refs.~ \cite{Cooper} and \cite{Berges} and introduce the CTP
formalism in our equation of motion \ by using the composition rule for
transition amplitudes along the time contour in the complex plane. This way
is cleaner notationally \ and has the advantage that all the functional
formalism of the previous section may be taken with the only difference of
path ordering according to the complex \ time contour $C_{ctp}$ in the time
integrations.

The two-point functions are decomposed as

\begin{equation}
G_{ij}^{ab}(t,t^{\prime })=\theta _{ctp}(t,t^{\prime
})G_{ij}^{ab>}(t,t^{\prime })+\theta _{ctp}(t^{\prime
},t)G_{ij}^{ab<}(t,t^{\prime }),
\end{equation}

\noindent where

\begin{eqnarray}
\hbar G_{ij}^{ab>}(t,t^{\prime }) &=&\left\langle \varphi_{i}^{a}(t)%
\varphi_{j}^{b}(t^{\prime })\right\rangle, \\
\hbar G_{ij}^{ab<}(t,t^{\prime }) &=&\left\langle \varphi%
_{i}^{b}(t^{\prime })\varphi_{j}^{a}(t)\right\rangle,
\end{eqnarray}

\noindent with ${\varphi}_{i}$ being the fluctuation field and
$\theta _{ctp}(t-t^{\prime })$ being the CTP complex contour
ordered theta function defined by

\begin{equation}
\theta _{ctp}(t,t^{\prime })=\left\{
\begin{array}{c}
\theta (t,t^{\prime })\mathrm{\ for}\quad t\quad \mathrm{\ and}\quad
t^{\prime }\quad \mathrm{\ both}\quad \mathrm{on}\quad C^{+} \\
\theta (t^{\prime },t)\mathrm{\ for}\quad t\quad \mathrm{\ and}\quad
t^{\prime }\quad \mathrm{\ both}\quad \mathrm{on}\quad C^{-} \\
1\;\mathrm{for}\quad t\quad \mathrm{\ on}\quad C^{-}\quad \mathrm{\ and}%
\quad \quad t^{\prime }\quad \mathrm{\ on}\quad C^{+} \\
0\ \mathrm{\ for}\quad t\quad \mathrm{\ on}\quad C^{+}\quad \mathrm{\ and}%
\quad \quad t^{\prime }\quad \mathrm{\ on}\quad C^{-}
\end{array}
\right.
\end{equation}

\begin{center}
With these definitions \ the matrix indices are not required. When
integrating over the second half $C^{-}$, we have to multiply \ by a \
negative sign to take into account the \ opposite direction of integration.
\end{center}

To show \ explicitly that \ \ the prescription for the CTP integration \
explained above \ does lead to a well-posed initial value problem \ with
causal equations, let us explicitly consider the \ integral in Eq. (\ref
{fluc}). The integrand has the CTP ordered form

\begin{eqnarray}
\Sigma (t,t^{\prime \prime })G(t^{\prime \prime },t^{\prime }) &=&\theta
_{ctp}(t,t^{\prime \prime })\theta _{ctp}(t^{\prime \prime },t^{\prime
})\Sigma ^{>}(t,t^{\prime \prime })G^{>}(t^{\prime \prime },t^{\prime
})+\theta _{ctp}(t,t^{\prime \prime })\theta _{ctp}(t^{\prime },t^{\prime
\prime })\Sigma ^{>}(t,t^{\prime \prime })G^{<}(t^{\prime \prime },t^{\prime
}) \\
&&\theta _{ctp}(t^{\prime \prime },t)\theta _{ctp}(t^{\prime \prime
},t^{\prime })\Sigma ^{<}(t,t^{\prime \prime })G^{>}(t^{\prime \prime
},t^{\prime })+\theta _{ctp}(t^{\prime \prime },t)\theta _{ctp}(t^{\prime
},t^{\prime \prime })\Sigma ^{<}(t,t^{\prime \prime })G^{<}(t^{\prime \prime
},t^{\prime }),  \notag
\end{eqnarray}
where we have omitted the indices because they are not relevant for the
discussion. Using the rule for CTP\ contour integration \ we get

\begin{eqnarray}
\int dt^{\prime \prime }\Sigma (t,t^{\prime \prime })G(t^{\prime \prime
},t^{\prime }) &=&\int_{0}^{t}dt^{\prime \prime }\left( \theta (t^{\prime
\prime },t^{\prime })\Sigma ^{>}(t,t^{\prime\prime })G^{>}(t^{\prime \prime
},t^{\prime })+\theta (t^{\prime\prime },t^{\prime \prime })\Sigma
^{>}(t,t^{\prime })G^{<}(t^{\prime \prime },t^{\prime })\right)  \notag \\
&&+\int_{t}^{\infty }dt^{\prime \prime }\left( \theta (t^{\prime \prime
},t^{\prime })\Sigma ^{<}(t,t^{\prime\prime })G^{>}(t^{\prime \prime
},t^{\prime })+\theta (t^{\prime },t^{\prime \prime })\Sigma
^{<}(t,t^{\prime\prime })G^{<}(t^{\prime \prime },t^{\prime })\right)  \notag
\\
&&-\int_{0}^{\infty }dt^{\prime \prime }\Sigma ^{<}(t,t^{\prime\prime
})G^{>}(t^{\prime \prime },t^{\prime }).
\end{eqnarray}
If $t>t^{\prime }$, we have

\begin{equation}
\int dt^{\prime \prime }\Sigma (t,t^{\prime \prime })G(t^{\prime \prime
},t^{\prime })=\int_{0}^{t}dt^{\prime \prime }(\Sigma ^{>}(t,t^{\prime
\prime})-\Sigma ^{<}(t,t^{\prime \prime}))G^{>}(t^{\prime \prime },t^{\prime
})-\int_{0}^{t^{\prime }}dt^{\prime \prime }\Sigma ^{>}(t,t^{\prime\prime
})(G^{>}(t^{\prime \prime },t^{\prime })-G^{<}(t^{\prime \prime },t^{\prime
})).
\end{equation}
On the other hand, if $t<t^{\prime }$

\begin{equation}
\int dt^{\prime \prime }\Sigma (t,t^{\prime \prime })G(t^{\prime \prime
},t^{\prime })=\int_{0}^{t}dt^{\prime \prime }(\Sigma ^{>}(t,t^{\prime\prime
})-\Sigma ^{<}(t,t^{\prime\prime }))G^{<}(t^{\prime \prime },t^{\prime
})-\int_{0}^{t^{\prime} }dt^{\prime \prime }\Sigma ^{>}(t,t^{\prime
\prime})(G^{>}(t^{\prime \prime },t^{\prime })-G^{<}(t^{\prime \prime
},t^{\prime })).
\end{equation}
The above equations \ are explicitly causal.

It is convenient to express the evolution equations in terms \ of two
independent \ two-point functions \ which can be associated to the
expectation values of the commutator and the anti-commutator of the fields.
We define, following \ Ref.~\cite{Berges} the functions

\begin{eqnarray}
G_{ij}^{(F)ab}(t,t^{\prime }) &=&\frac{1}{2}\left( G_{ij}^{ab>}(t,t^{\prime
})+G_{ij}^{ab<}(t,t^{\prime })\right) \\
G_{ij}^{(\rho )ab}(t,t^{\prime }) &=&i\left( G_{ij}^{ab>}(t,t^{\prime
})-G_{ij}^{ab<}(t,t^{\prime })\right) ,
\end{eqnarray}
where the $(F)$ \ functions \ are usually called statistical
propagators and the $(\rho )$, spectral functions. (See Ref.
\cite{Kadanoff}). With these definitions Eq.(\ref{fluc}) can be
rewritten as:

\begin{eqnarray}
h_{c}^{a}\hbar \partial _{t}G_{ij\;}^{\;(F)cb}(t,t^{\prime }) &=&-J\left(
G_{i+1j}^{(F)ab}(t,t^{\prime })+G_{i-1j}^{(F)ab}(t,t^{\prime })\right) +%
\frac{U}{\mathcal{N}}\left( \phi _{ic}(t)\phi
_{i}^{c}(t))G_{ij}^{(F)ab}(t,t^{\prime })\right) + \\
&&\frac{2U}{\mathcal{N}}\left( \phi _{i}^{a}(t)G_{ij}^{(F)cb}(t,t^{\prime
})\phi _{ic}(t)\right) +\int_{0}^{t}dt_{k}^{\prime \prime }\Sigma
_{ik}^{(\rho )ac}(t,t^{\prime \prime })G_{kj\;c}^{(F)\;b}(t^{\prime \prime
},t^{\prime })  \notag \\
&&-\int_{0}^{t^{\prime }}dt_{k}^{\prime \prime }\Sigma
_{ik}^{(F)ac}(t,t^{\prime \prime })G_{kjc}^{(\rho )b}(t^{\prime \prime
},t^{\prime }),  \notag \\
h_{c}^{a}\hbar \partial _{t}G_{ij\;}^{\;(\rho )cb}(t,t^{\prime })
&=&-J\left( G_{i+1j}^{(\rho )ab}(t,t^{\prime })+G_{i-1j}^{(\rho
)ab}(t,t^{\prime })\right) +\frac{U}{\mathcal{N}}\left( \phi _{ic}(t)\phi
_{i}^{c}(t))G_{ij}^{(\rho )ab}(t,t^{\prime }\right) + \\
&&\frac{2U}{\mathcal{N}}\left( \phi _{i}^{a}(t)G_{ij}^{(\rho
)cb}(t,t^{\prime })\phi _{ic}(t)\right) +\int_{t^{\prime
}}^{t}dt_{k}^{\prime \prime }\Sigma _{ik}^{(\rho )ac}(t,t^{\prime \prime
})G_{kjc}^{(\rho )b}(t^{\prime \prime },t^{\prime }).  \notag
\end{eqnarray}
In particular, we define the normal, $\rho $, and anomalous, $\ m$, spectral
and statistical functions as

\begin{eqnarray}
G_{ij}^{21(F)}(t,t^{\prime }) &\equiv &\rho _{ij}^{(F)}(t,t^{\prime })=\frac{%
1}{2}\left\langle \varphi_{i}^{\dagger }(t)\varphi%
_{j}(t^{\prime })+\varphi_{j}(t^{\prime })\varphi%
_{i}^{\dagger }(t)\right\rangle ,  \label{rof} \\
G_{ij}^{21(\rho )}(t,t^{\prime }) &\equiv &\rho _{ij}^{(\rho )}(t,t^{\prime
})=i\left\langle \varphi_{i}^{\dagger }(t)\varphi%
_{j}(t^{\prime })-\varphi_{j}(t^{\prime })\varphi%
_{i}^{\dagger }(t)\right\rangle , \\
G_{ij}^{11(F)}(t,t^{\prime }) &\equiv &m_{ij}^{(F)}(t,t^{\prime })=\frac{1}{2%
}\left\langle \varphi_{i}(t)\varphi_{j}(t^{\prime })+\tilde{%
\varphi}_{j}(t^{\prime })\varphi_{i}(t)\right\rangle , \\
G_{ij}^{11(\rho )}(t,t^{\prime }) &\equiv &m_{ij}^{(\rho )}(t,t^{\prime
})=i\left\langle \varphi_{i}(t)\varphi_{j}(t^{\prime })-%
\varphi_{j}(t^{\prime })\varphi_{i}(t)\right\rangle , \label{mro}
\end{eqnarray}

With these relations in place, we now proceed to derive the time evolution
equations for the mean field and \ the two-point functions \ from the CTP
2PI effective action for the Bose-Hubbard Model under the three
approximations described before.

\section{HFB Approximation}


As remarked in Section 2 the first order mean field approximation leads to a
DNLSE which includes only the contribution from the condensate. The HFB
equations go beyond the first order approximation and \ include \ the
leading order contribution of $\ \Gamma _{2}$. They describe the coupled
dynamics of condensate and non-condensate \ atoms\ which arise from the most
important scattering processes which are direct, exchange and pair
excitations. The basic damping mechanism present in the HFB approximation
are Landau and Beliaev damping associated with the decay of an elementary
excitation into a pair of excitations in the presence of condensate atoms (
\cite{Giorgini,HuPavon}). However, these kinds of damping \footnote{%
we make a distinction between the meaning of the words `damping' and
`dissipation', the former referring simply to the phenomenological decay of
some function, the latter with theoretical meaning, e.g., in the Boltzmann
sense} found in the HFB approximation (due to phase mixing, as in the Vlasov
equation \cite{Balescu}) are different in nature from the collisional
dissipation (as in the Boltzmann equation) responsible for thermalization
processes. Multiple scattering processes are neglected in this
approximation. We expect the HFB equations to give a good description of the
dynamics in the collisionless regime when interparticle collisions play a
minor role.

The leading order contribution of $\ \Gamma _{2}$ \ is represented by \ the
\emph{double-bubble} diagram. Its contribution to $\Gamma _{2}$ is $\phi $
independent and has an analytic expression of the form
\begin{equation}
\Gamma _{2}^{(1)}[G]=-\frac{U}{4\mathcal{N}}\int dt_{i}\;\left(
G_{iia}^{\;a}(t,t)G_{iib}^{\;b}(t,t)+2G_{iiab}(t,t)G_{ii}^{ab}(t,t)\right),
\label{ga1}
\end{equation}
the factor of two arises because \ the direct and exchange terms are
identical.

Using the first order expression for $\Gamma _{2}$ in Eqs. ($\ref{cond}$)
and (\ref{fluc}) yields the following equations of motion.

\begin{eqnarray}
h_{b}^{a}\hbar \partial _{t}\phi _{i}^{b}(t)&=&\Im_{HFB}^{\phi}, \\
\Im_{HFB}^{\phi}&\equiv&-J \left(\phi _{i+1}^{a}(t)+\phi
_{i-1}^{a}(t)\right)+\frac{U}{\mathcal{N}}\left(\phi _{id}(t)\phi
_{i}^{d}(t)+G_{ii\;d}^{\;d}(t,t)\right)\phi _{i}^{a}(t)+\frac{2U}{\mathcal{N}%
}\left(\phi _{i}^{b}(t)G_{iiba}^{\;\ }(t,t)\right),  \notag  \label{defhfbp}
\end{eqnarray}

\begin{eqnarray}
h_{c}^{a}\partial _{t}G_{ij}^{cb\;\ }(t,t^{\prime }) &=&\Im_{HFB}^{G}, \\
\Im_{HFB}^{G}&\equiv&-J(G_{i+1j}^{ab}(t,t^{\prime })+
G_{i-1j}^{ab}(t,t^{\prime }))+\frac{U}{\mathcal{N}}(\phi _{id}(t)\phi
_{i}^{d}(t)+G_{ii\;d}^{\;d}(t,t))G_{ij}^{ab}(t,t^{\prime })+  \notag \\
&&\frac{2U}{\mathcal{N}}(\phi _{i}^{a}(t)\phi _{ic}(t)G_{ij}^{cb\;\
}(t,t^{\prime })+G_{ii\;d}^{\;a}(t,t)G_{ij}^{db}(t,t^{\prime }))+i\delta
^{ab}\delta _{ij}\delta _{C}(t-t^{\prime }).  \notag  \label{defhfbG}
\end{eqnarray}

In terms of the spectral and statistical functions, \ Eqs. (\ref{rof}) to (%
\ref{mro}), and setting $\mathcal{N}=2$, the above equations take the form

\begin{eqnarray}
i\hbar \partial _{t}\phi _{i}(t)&=&-J(\phi _{i+1}(t)+\phi _{i-1}(t))+U(|\phi
_{i}(t)|^{2}+2\rho _{ii}^{(F)}(t,t))\phi _{i}(t)+Um_{ii}^{(F)}(t,t)\phi
_{i}^{\ast }(t),  \label{hfbc}
\end{eqnarray}
\begin{eqnarray}
-i\hbar \frac{\partial }{\partial t}\rho _{ij}^{(F)}(t,t^{\prime }) &=&%
\mathbf{L}_{ik}(t)\rho _{kj}^{(F)}(t,t^{\prime })+\mathbf{M}_{ik}^{*}(t)
m_{kj}^{(F)}(t,t^{\prime }), \\
-i\hbar \frac{\partial }{\partial t}\rho _{ij}^{(\rho )}(t,t^{\prime }) &=&%
\mathbf{L}_{ik}(t)\rho _{kj}^{(\rho)}(t,t^{\prime })+\mathbf{M}_{ik}^{*}(t)
m_{kj}^{(\rho)}(t,t^{\prime }), \\
i\hbar \frac{\partial }{\partial t}m_{ij}^{(F)}(t,t^{\prime }) &=&\mathbf{L}%
_{ik}(t)m _{kj}^{(F)}(t,t^{\prime })+\mathbf{M}_{ik}(t)
\rho_{kj}^{(F)}(t,t^{\prime }), \\
i\hbar \frac{\partial }{\partial t}m_{ij}^{(\rho )}(t,t^{\prime }) &=&%
\mathbf{L}_{ik}(t)m _{kj}^{(\rho)}(t,t^{\prime })+\mathbf{M}_{ik}(t)
\rho_{kj}^{(\rho)}(t,t^{\prime }),  \label{hfbmm}
\end{eqnarray}

with

\begin{eqnarray}
\mathbf{L}_{ij}(t)&=&-J(\delta _{i+1j}+\delta _{i-1j})+2U\delta _{ij}\left(
|\phi _{i}(t)|^{2}+\rho^{F}_{ii}(t,t)\right), \\
\mathbf{M}_{ij}(t)&=&U\delta _{ij}\left( \phi
_{i}(t)^{2}+m^{F}_{ii}(t,t)\right).
\end{eqnarray}

The time dependent HFB \ equations \ are a closed set of
self-consistent equations \ that describe the coupled dynamics of
the condensate \ and non-condensate \ components of a Bose gas. It
can be checked \ that they preserve important conservation laws
such as the number of particles and energy. The conservation
properties of the HFB equations can also be understood by the fact
that these equations can also be derived using Gaussian
variational methods \cite{Griffin}. These methods always yield a
classical Hamiltonian dynamics which guarantees probability
conservation. Because they are local in time they can be decoupled
by a mode decomposition. (See Appendix 1 for details).

\section{Second order expansion}

\subsection{Equations of Motion}

\subsubsection{Full second order}

The second order contribution to\ $\Gamma _{2}$ is described in terms of the
\emph{setting-sun} Fig. (2b) and the \emph{basket-ball} Fig. (2c) diagrams.
The \emph{basket-ball} diagram is independent of the mean- field \ and is
constructed with only \ quartic vertices. The \emph{setting-sun} diagram
depends on $\phi $ and contains only three-point vertices. \ The second
order $\Gamma _{2}^{(2)}$ effective action can be written as

\begin{eqnarray}
&&\Gamma _{2}^{(2)}[\phi ,G]=  \notag \\
&&i{\left(\frac{U}{\mathcal{N}}\right)}^{2}\int dt_{i}dt_{j}\phi
_{ib}(t)\phi _{jb^{\prime }}(t^{\prime })\left( G_{ij}^{bb^{\prime
}}(t,t^{\prime })G_{ijdd^{\prime }}(t,t^{\prime })G_{ij}^{dd^{\prime
}}(t,t^{\prime })+2G_{ij}^{bd^{\prime }}(t,t^{\prime })G_{ijdd^{\prime
}}(t,t^{\prime })G_{ij}^{db^{\prime }}(t,t^{\prime })\right) + \\
&&i\left(\frac{U}{2 \mathcal{N}}\right)^{2}\int dt_{i}dt_{j}^{\prime
}\;\left( G_{ijbb^{\prime }}(t,t^{\prime })G_{ij}^{bb^{\prime }}(t,t^{\prime
})G_{ijdd^{\prime }}^{{}}(t,t^{\prime })G_{ij}^{dd^{\prime }}(t,t^{\prime
})+2G_{ijbb^{\prime }}^{{}}(t,t^{\prime })G_{ij}^{bd^{\prime }}(t,t^{\prime
})G_{ijdd^{\prime }}^{{}}(t,t^{\prime })G_{ij}^{db^{\prime }}(t,t^{\prime
})\right).  \notag  \label{ga2}
\end{eqnarray}

To simplify the notation, let us introduce the following definitions \cite
{Berges}:

\begin{eqnarray}
\Pi _{ij}(t,t^{\prime }) &=&-\frac{1}{2}G_{ijab}(t,t^{\prime
})G_{ij}^{ab}(t,t^{\prime }), \\
\Xi _{ijab}^{\;\;\;}(t,t^{\prime }) &=&-D(t,t^{\prime
})G_{ijab}^{\;\;\;\;}(t,t^{\prime }), \\
D(t,t^{\prime }) &=&\phi _{ib}(t)\phi _{ja}(t^{\prime
})G_{ij\;\;}^{ba}(t,t^{\prime })-\Pi _{ij}^{\;\;}(t,t^{\prime }), \\
\overline{\Lambda }_{ij\;a}^{\;b}(t,t^{\prime }) &=&-G_{ij}^{cb}(t,t^{\prime
})G_{ijca}^{\;}(t,t^{\prime }) , \\
\Lambda _{ij\;a}^{\;b}(t,t^{\prime }) &=&-G_{ij\;}^{\;bc}(t,t^{\prime
})G_{ijac}^{\;\;\;^{\;}}(t,t^{\prime }), \\
\Theta _{ijac}^{\;\;\;}(t,t^{\prime }) &=&-(\phi _{id}(t)\phi
_{jb}(t^{\prime })+G_{ijdb}(t,t^{\prime }))G_{ij}^{ab}(t^{\prime
},t)G_{ij}^{dc}(t,t^{\prime })+\Xi _{ijac}^{\;\;\;}(t,t^{\prime }).
\end{eqnarray}
With the above definitions we find from Eqs. (\ref{cond}) and \ (\ref{fluc})
the following equations of motion:

\
\begin{eqnarray}
h_{b}^{a}\hbar \partial _{t}\phi _{i}^{b}(t) &=&\Im _{HFB}^{\phi }+i\left(
\frac{2U}{\mathcal{N}}\right) ^{2}\int dt_{j}^{\prime }\;\phi
_{jb}(t^{\prime })\left( \Pi _{ij}(t,t^{\prime })G_{ji\;\;}^{ba}(t^{\prime
},t)+\overline{\Lambda }_{ij\;c}^{\;b}(t,t^{\prime })G_{ij}^{ac}(t,t^{\prime
})\right) ,  \label{mean} \\
h_{c}^{a}\hbar \partial _{t}G_{ij}^{cb\;\ }(t,t^{\prime }) &=&\Im
_{HFB}^{G}+i\left( \frac{2U}{\mathcal{N}}\right) ^{2}\phi _{i}^{a}(t)\int
dt_{k}^{\prime \prime }\phi _{kc}(t^{\prime \prime })\left( \Pi
_{ik}(t,t^{\prime \prime })G_{kj}^{cb}(t^{\prime \prime },t^{\prime })+%
\overline{\Lambda }_{ik\;d}^{\;c}(t,t^{\prime \prime })G_{kj}^{db}(t^{\prime
\prime },t^{\prime })\right)   \notag \\
&&+i\left( \frac{2U}{\mathcal{N}}\right) ^{2}\int dt^{\prime \prime }\left(
\Theta _{ik\;d}^{\;a\;\;\;}(t,t^{\prime \prime })+\Lambda
_{ik\;}^{ca}(t,t^{\prime \prime })\phi _{ic}^{{}}(t)\phi _{kd}(t^{\prime
\prime })\right) G_{kj}^{db\ }(t^{\prime \prime },t^{\prime }),
\end{eqnarray}
where $\Im _{HFB}^{\phi }$ and $\Im _{HFB}^{G}$ are defined in Eqs. (\ref
{defhfbp}) and (\ref{defhfbG}). For  explicit expressions in terms of $\rho
^{(F,\rho )\text{ \ }}$ and $m^{(F,\rho )\text{ \ }}$ see Appendix 2.

\subsubsection{ 2PI-1/$\mathcal{N}$ expansion}

The 2PI effective action is a singlet under O($\mathcal{N}$) rotations. It
can be shown that all graphs contained in an O($\mathcal{N}$) \ expansion
can be built from the irreducible invariants\cite{Berges}: $\phi ^{2},$ $%
tr(G^{n})$ and $tr(\phi \phi G^{n})$ , with $n<\mathcal{N}$. \ The factors
of $\mathcal{N}$ in a single graph contributing to the same 1/$\mathcal{N}$
expansion have then two origins: a factor of $\mathcal{N}$ from each
irreducible invariant and a factor of 1/$\mathcal{N}$ from each vertex. The
leading order large $\mathcal{N}$ approximation \ scales proportional to $%
\mathcal{N}$, the next to leading order (NLO) \ contributions are of order 1
and so on. At leading order only the first term of Eq. (\ref{ga1})
contributes.\ At the next to leading order level, if we truncate up to
second order in the coupling strength, the \emph{double-bubble} is totally
included but only certain parts of the \emph{setting-sun} and \emph{%
basket-ball} diagrams are: the first term in both of the integrals of Eq. (%
\ref{ga2}),

\begin{eqnarray}
\Gamma _{2}^{(2) 1/\mathcal{N}}[\phi ,G]&=& i{\left(\frac{U}{\mathcal{N}}%
\right)}^{2}\int dt_{i}dt_{j}\phi _{ib}(t)\phi _{jb^{\prime }}(t^{\prime
})\left( G_{ij}^{bb^{\prime }}(t,t^{\prime })G_{ijdd^{\prime }}(t,t^{\prime
})G_{ij}^{dd^{\prime }}(t,t^{\prime })\right)+  \notag \\
&&i\left(\frac{U}{2 \mathcal{N}}\right)^{2}\int dt_{i}dt_{j}^{\prime
}\;\left( G_{ijbb^{\prime }}(t,t^{\prime })G_{ij}^{bb^{\prime }}(t,t^{\prime
})G_{ijdd^{\prime }}^{{}}(t,t^{\prime })G_{ij}^{dd^{\prime }}(t,t^{\prime
})\right).  \label{ga2ln}
\end{eqnarray}

\noindent The equations of motion under this approximation are the ones
obtained for the full second order expansion but with $\Lambda =\overline{%
\Lambda }=0$, and $\Theta =\Xi .$

\bigskip

In Appendix 2 we explicitly write the equations of motion in terms of the
spectral and statistical functions. We end this section by emphasizing that
the only approximation introduced in the derivation of the equations of
motion presented here is the truncation up to second order in the
interaction strength. These equations depict the nonlinear and non-Markovian
quantum dynamics, which we consider as the primary distinguishing features
of this work. It supersedes what the second order kinetic theories currently
presented can do, their going beyond the HFB approximation notwithstanding.
For example Ref. \cite{Holland} presents a kinetic theory approach that
includes binary interactions to second order in the interaction potential
but uses the Markovian approximation. In Ref.~\cite{Holland2} the authors
gave a non-Markovian generalization to the quantum kinetic theory derived by
Walser \emph{et. al.}\cite{Walser} by including memory effects. However in
that work symmetry breaking fields,$\phi$ and anomalous fluctuations $m$ are
neglected.

\subsection{Conservation Laws}

For a closed (isolated) system the mean total number of particles N and
energy are conserved quantities as they are the constants of motion for the
dynamical equations.

Particle number conservation is a consequence of the invariance of the
Hamiltonian under a global phase change. The mean total number of particles
is given by

\begin{eqnarray}
\left\langle \hat{N}\right\rangle &=&\sum_{i}\left\langle \hat{\Phi}%
_{i\;}^{\dagger }\hat{\Phi}_{i}\right\rangle =\sum_{i}\left( |\phi
_{i}|^{2}+\rho _{ii}^{(F)}-\frac{1}{2}\right) \\
&=&N.  \notag
\end{eqnarray}
The kinetic equation for $N$ is then

\begin{eqnarray}
\frac{d}{dt}\left\langle \hat{N}(t)\right\rangle &=&\sum_{i}2Re \left( \phi
_{i}(t)\frac{\partial }{\partial t}\phi _{i}^{\ast }(t)\right)
+\lim_{t->t^{\prime }}\left( \frac{\partial }{\partial t}\rho
_{ii}^{(F)}(t,t^{\prime })+\frac{\partial }{\partial t^{\prime }}\rho
_{ii}^{\ast (F)}(t^{\prime },t)\right) \\
&=&0.  \notag
\end{eqnarray}

All three approximations we have considered, namely, HFB, 1/$\mathcal{N}$
expansion and full second order expansion, conserve particle number. This
can be shown by plugging in the kinetic equation of $\left\langle \hat{N}%
(t)\right\rangle $ the equation of motion for the mean field and the normal
\ statistical \ propagator and cancelling terms. It is important to note
that even though total population is always conserved there is always a
transfer of population between condensate and non condensate atoms.

While number conservation can be proved explicitly, to prove total energy
conservation is not obvious as the Hamiltonian cannot be represented as a
linear combination of the relevant operators. It is clear that the exact
solution of a closed system is unitary in time and hence disallows any
dissipation. However, the introduction of approximation schemes that
truncate the infinite hierarchy of correlation functions at some finite
order with causal boundary conditions may introduce dissipation \cite{mea}.

To discuss energy conservation we can use the phi-derivable criteria \cite
{Baym} which states that nonequilibrium approximations in which the self
energy $\Sigma$ is of the form $\delta\Phi/\delta G$, with $\Phi$ a
functional of $G$, conserve particle number, energy and momentum. All the
approximations we consider in this paper are phi derivable and thus they
obey energy, particle number and momentum conservation laws. For HFB, $\Phi=
\Gamma_{2}^{(1)}$, for the full second order expansion, $\Phi=
\Gamma_{2}^{(1)}+\Gamma_{2}^{(2)}$ and for the second order next to leading
order $1/\mathcal{N}$ expansion, $\Phi= \Gamma_{2}^{(1)}+\Gamma_{2}^{(2)1/%
\mathcal{N}}$. See Eqs. (\ref{pder}), (\ref{ga1}), (\ref{ga2}) and (\ref
{ga2ln}). For a detailed discussion of the complete next to leading order $1/%
\mathcal{N}$ expansion see Refs. \cite{Cooper,Berges} and references therein.

\subsection{Zero mode fluctuactions}

The spectrum of fluctuations above the condensate includes a zero mode. This
mode is the Goldstone boson associated with the breaking of global phase
invariance by the condensate. It is analogous to the collective modes which
arise in the spectrum of fluctuations around a bubble \cite{R. Rajaraman} .
The zero mode is essentially non perturbative. In linearized theory, it
introduces an artificial infrared divergence in low dimensional models. For
this reason linearized theory is actually improved if the contribution from
this mode is neglected all together \cite{J. Andersen}. A different way to
deal with the zero mode has been proposed by Gardiner \cite{Gardiner} and
Morgan \cite{Morgan}. Here the theory is written in terms of operators which
exchange particles between zero and nonzero modes, conserving the total
particle number, and one further operator which changes total particle
number. The contribution from the zero mode is then subtracted by expressing
the normal and anomalous densities in terms of the former alone. However,
from a physical point of view the zero mode exists and is quantum in nature.
We may think of it as the limit of Gardiner and Zoller's \cite{Gardiner2}
"condensate band "when the width of the band shrinks to zero. There are both
fundamental and practical reasons why isolating and subtracting the zero
mode is not as compelling in our case as in the problems discussed by
Gardiner and Morgan. In the problem we discuss, the initial state is a
coherent state rather than a proper state of the total particle number. As
the total particle number is not very high quantum fluctuations in the total
particle number are real, and non-negligible. Discarding these fluctuations
would spoil the integrity of the formalism. Also, because the 2PI formalism
goes beyond the linearized approximation, the zero mode does not have the
impact it has in the linearized formalism and it is not clear that
subtracting it necessarily leads to a better approximation. Therefore, in
this paper we shall not attempt to isolate the contributions from the zero
mode. A full non-perturbative treatment in the future is certainly desirable.

\section{Numerical Implementation}

\subsection{Exact Solution}

The fully quantal solution was found by evolving in time the initial state
with the Bose-Hubbard Hamiltonian given by Eq. (\ref{BHH}),so that $%
|\varphi(t)\rangle=e^{-\frac{i}{\hbar} \hat{H}t}|\varphi(0)\rangle$ with $%
|\varphi(0)\rangle=e^{-N/2}e^{\sqrt{N}\hat{\Phi}_{0}^{\dagger
}}\left|0\right\rangle_{0}\prod_{i\neq0}|0\rangle _{i}$. To do the
numerical calculations we partitioned the Hilbert space in
subspaces with fixed number of atoms and propagated independently
the projections of the initial state on the respective subspaces.
A subspace with $N_{n}$ number of atoms and $I$ number of wells is
spanned by $\frac{(N_{n}+I-1)!}{N_{n}!(I-1)!}$ states. This
procedure could be done because the Hamiltonian commutes with the
number operator $\sum_{i}\hat{\Phi_{i}}^{\dagger }\hat{\Phi_{i}}$,
and thus during the dynamics the different subspaces never get
mixed. The number of subspaces used for the numerical evolution
were such that no change in plots of the dynamical observables was
detected by adding another subspace. Generally for N atoms in the
initial state, this condition was achieved by including the
subspaces between $N-4\sqrt{N}$ and $N+4\sqrt{N}$ atoms.

\subsection{Numerical Algorithm for the approximated solution}

The time evolution equations obtained in Sec. 7 are nonlinear
integro-differential equations. Though the equations are very
complicated, they can be solved on a computer. The important point
to note is that all equations are casual in time, and all
quantities at some later time $t_{f}$ can be obtained by
integration over the explicitly known functions for times $t\leq
t_{f}$.

For the numerical solution we employed a time discretization $t=na_{t},$ $%
t^{\prime }=ma_{t},$ and took the advantage that, due to the presence of the
lattice the spatial dimension is discrete (indices $i$ and $j$). The
discretized equations for the time evolution of the matrices $\rho
_{ijnm}^{(F,\rho )},$ $m_{ijnm}^{(F,\rho )}$ and $\phi _{in}$ advance time
step wise in the $n$-direction for fixed $m$. Due to the symmetries of the
matrices only half of the $(n,m)$ matrices have to be computed and the
values $\rho _{ijnn}^{\rho }=-i,$ $m_{ijnn}^{\rho }=0$ are fixed for all
time due to the bosonic commutation relations. As initial conditions one
specifies $\rho _{ij00}^{(F,\rho )},$ $m_{ij00}^{(F,\rho )}$ and $\phi _{i0.}
$

To ensure that the discretized equations retain the conservation properties
present in the continuous ones one has to be very careful in the evolution
of the diagonal terms of $\rho _{iinn}^{(F)}$ and take the limit $%
m\rightarrow n$ in a proper way:

\begin{eqnarray}
\rho _{ijn+1n+1}^{(F,\rho )}-\rho _{ijnn}^{(F,\rho )} &=&\left( \rho
_{ijn+1n}^{(F,\rho )}-\rho _{ijnn}^{(F,\rho )}\right) \pm \left( \rho
_{jin+1n}^{\ast (F,\rho )}-\rho _{jinn}^{\ast (F,\rho )}\right) , \\
m_{ijn+1n+1}^{(F,\rho )}-m_{ijnn}^{(F,\rho )} &=&\left( m_{ijn+1n}^{(F,\rho
)}-m_{ijnn}^{(F,\rho )}\right) \pm \left( m_{jin+1n}^{(F,\rho
)}-m_{jinn}^{(F,\rho )}\right) ,
\end{eqnarray}
with the positive sign for the statistical propagators, $(F)^{\prime }s,$
and negative for the spectral ones,\ $(\rho )^{\prime }s.$ We use the fourth
order Runge-Kutta algorithm to propagate the local part of the equations and
a regular one step Euler method to iterate the non local parts. For the
integrals we use the standard trapezoidal rule. Starting with $n=1$, for the
time step $n+1$ one computes successively all entries with $m=0......,n,n+1$
from known functions evaluated at previous times.

The time step $a_{t}$ was chosen small enough so that convergence was
observed, that is, further decreasing it did not change the results. The
greater the parameter $UN/J$ the smaller the time step required. The main
numerical limitation \ of the 2PI approximation is set by the time
integrals, which make the numerical calculations time and memory consuming.
However, within a typical numerical precision it was typically not necessary
to keep all the past of the two point functions in the memory. A
characteristic time, after which the influence of the early time in the late
time behavior is given by the inverse damping rate. This time is described
by the exponential damping of the two-point correlator at time t with the
initial time\cite{Berges}. In our numerics we extended the length of the
employed time interval until the results did not depend on it. In general,
it was less than the inverse damping rate. We used for the calculations a
single PII 400 MH workstation with 260 Mb of memory. For a typical run 1-2
days of computational time were required.

\subsection{ Initial conditions and parameters}

To model the patterned \ loading \ the initial conditions assumed for the
numerical solutions were $\phi _{i0}=N\delta _{i0}$,$\;\rho _{ij00}^{(F)}=%
\frac{1}{2}\delta _{ij}$,$\;\rho _{ij00}^{(\rho )}=-i\delta _{ij}$ $\ $and$%
\;m_{ij00}^{(F)}=m_{ij00}^{(\rho )}=0.$ They correspond to an initial
coherent state with $N$ atoms in the initial populated well.

To study the kinetic energy dominated regime we chose for the simulations
three different sets of parameters: The first set was chosen to be in the
very weak interacting regime, $I=3,N=6,J=1$ and $U/J=1/30$. With this choice
we wanted to show the validity of a mean field approach to describe this
regime and the corrections introduced by the higher order approximations.
The second set of parameters were $I=3,N=8,J=1$ and $U/J=1/3$. In this
regime the kinetic energy is big enough to allow tunneling but the effect of
the interactions are crucial in the dynamics. Comparing with the exact
solution we could show the break-down of the mean field approximation.

At the mean field level (using the DNLSE) for a given number of wells, the
only  relevant parameter for describing the dynamics of the system is the
ratio $UN/J$. For a fixed $UN/J$ the mean field dynamics is independent of
the number of atoms in consideration. This is not the case in the exact
solution where both $UN/J$ and $N$ are important. As $N$ is increased the
bigger the population in the initial coherent matter field and therefore we
expect a better agreement of the truncated theories with the exact solution.
To study the dependence of the dynamics on the total number of atoms, the
third set of parameters in our solutions were chosen to be $I=2,J=1/2$ and
fixed $NU/J=4$ but we changed the number of atoms from $20$ to $80$. To
increase the number of atoms in the calculations we had to reduce the number
of wells to two due to the fact that the dimension of the Hilbert space
scales very badly with $N$ and $I$.

\section{Results and Discussions}

\begin{figure}[tbh]
\begin{center}
\leavevmode {\includegraphics[width=7.2 in]{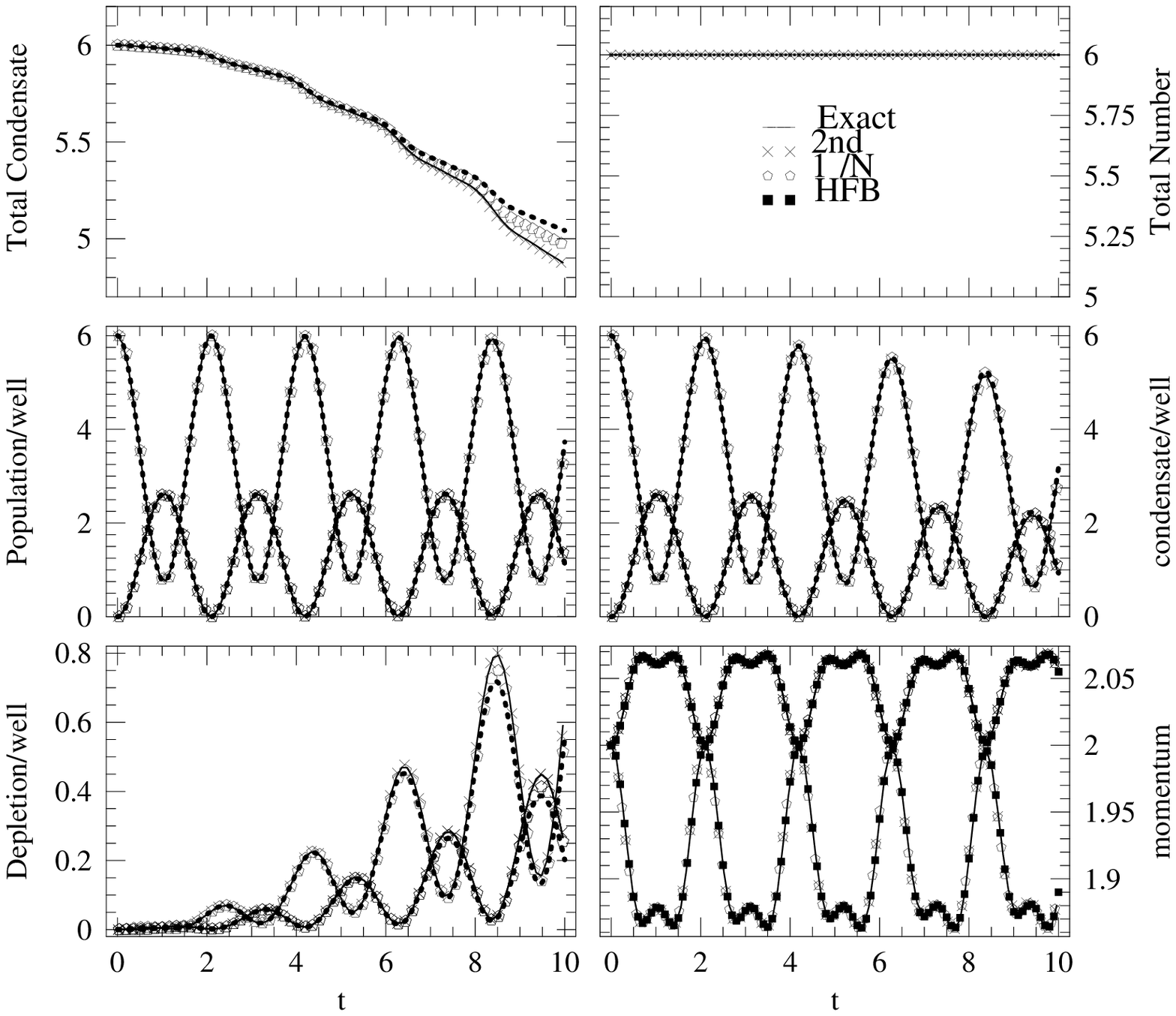}}
\end{center}
\caption{Comparisons between the exact solution(solid line), the HFB
approximation (boxes), the second order large $\mathcal{N}$ approximation
(pentagons) and the full 2PI second order approximation(crosses) for the
very weak interacting regime. The parameters used were $I=3,N=6,J=1$ and $%
U/J=1/30$. The time is given in units of $\hbar /J$. \ In the
plots where the population, condensate and depletion per well are
depicted the top curves correspond to the initially populated well
solutions and the lower to the initially empty wells. Notice the
different scale used in the  depletion plot. In the momentum
distribution plot the upper curve correspond to the $k=\pm
2\protect\pi /3$ intensities and the lower one to the $k=0$
quasi-momentum intensity. } \label{Fig3}
\end{figure}

\begin{figure}[tbh]
\begin{center}
\leavevmode {\includegraphics[width=8.2 in]{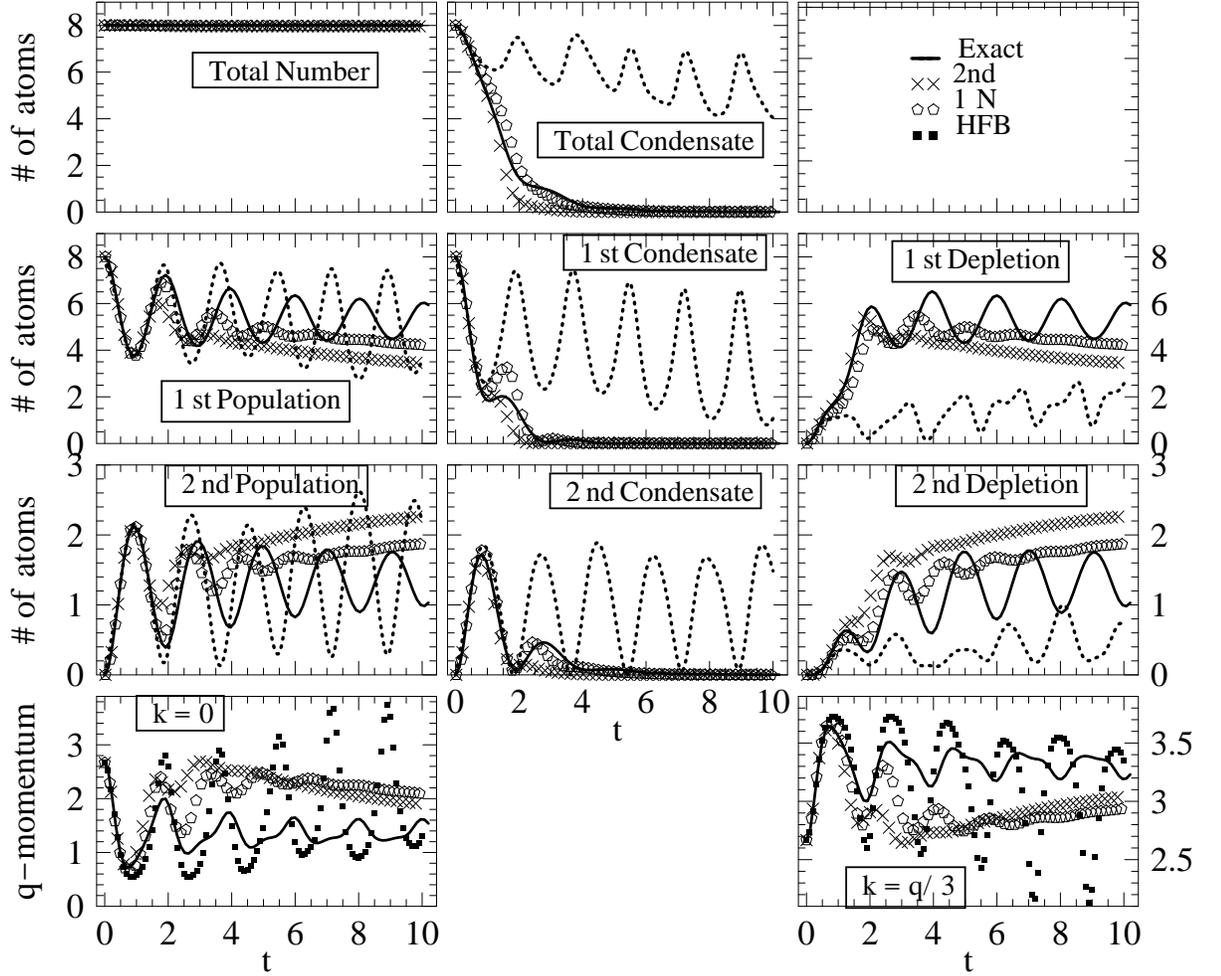}}
\end{center}
\caption{Comparisons for the case $I=3,N=8,J=1$ and $U/J=1/3$. The time is
given in units of $\hbar/J$.In the plots the abbreviation 1st is used for
the initially occupied well and 2nd for the initially empty wells. In the
quasi-momentum plots $q=2\protect\pi/a$ is the reciprocal lattice vector
with $a$ the lattice spacing. }
\label{Fig4}
\end{figure}

\begin{figure}[tbh]
\begin{center}
\leavevmode {\includegraphics[width=7.2 in]{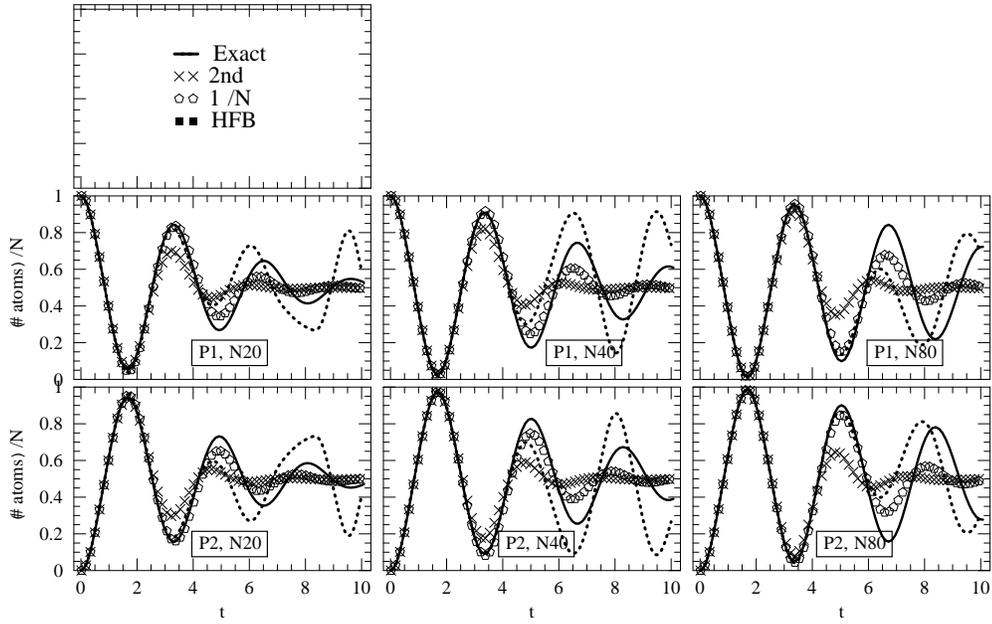}}
\end{center}
\caption{Comparison between the evolution of the atomic population per well
for $I=2,J=1/2$, $NU/J=4$ and $N=20,40$ and $80$. Time is in units of $\hbar
/J$. In the plots P1 states for the atomic population in the initially
populated wells and P2 for the population in the initially empty wells. The
number of atoms N is explicitly shown in each panel. To make easier the
comparisons all the plots are scaled by N such that the number of atoms is
always normalized to one. }
\label{Fig5a}
\end{figure}

\begin{figure}[tbh]
\begin{center}
\leavevmode {\includegraphics[width=7.2 in]{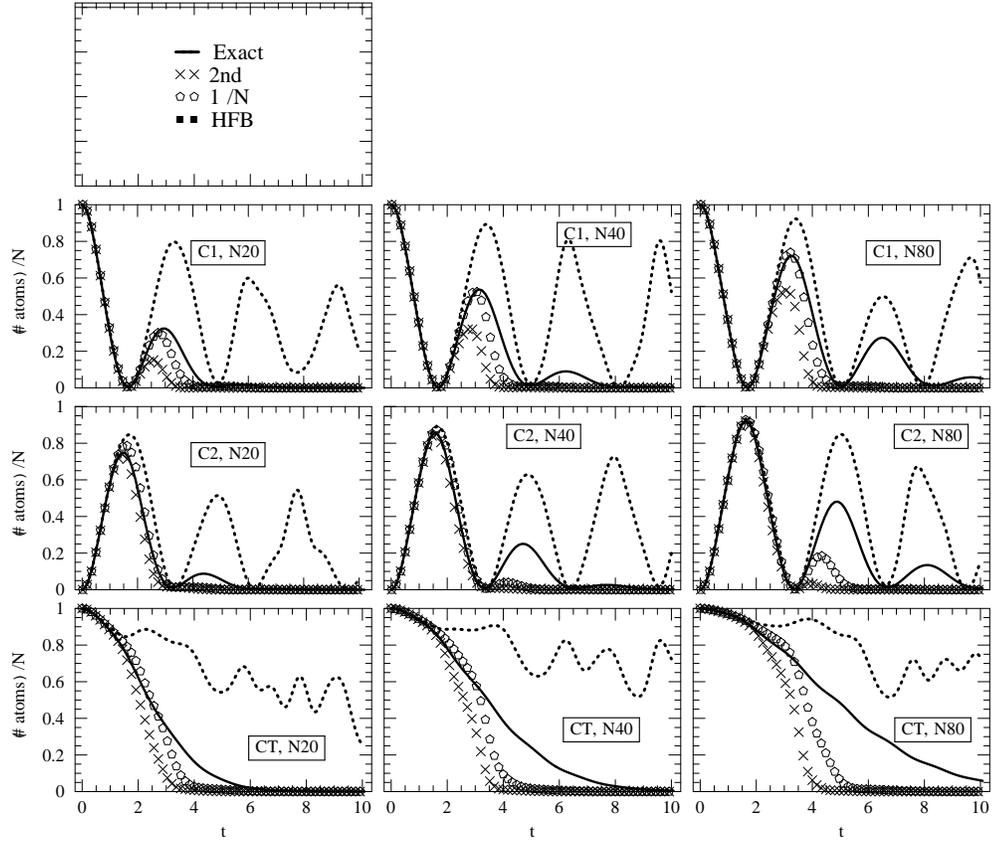}}
\end{center}
\caption{ Time evolution of the condensate population per well and the total
condensate population. Same paremeters than Fig.~5. Time is in units of $%
\hbar/J$. In the plots C1 states for the condensate population in the
initially populated well, C2 for the condensate in the initially empty one
and CT for the total condensate population. All the plots are scaled to have
the total number of atoms set to one for all the different N cases. }
\label{Fig5b}
\end{figure}

\begin{figure}[tbh]
\begin{center}
\leavevmode {\includegraphics[width=7.2 in]{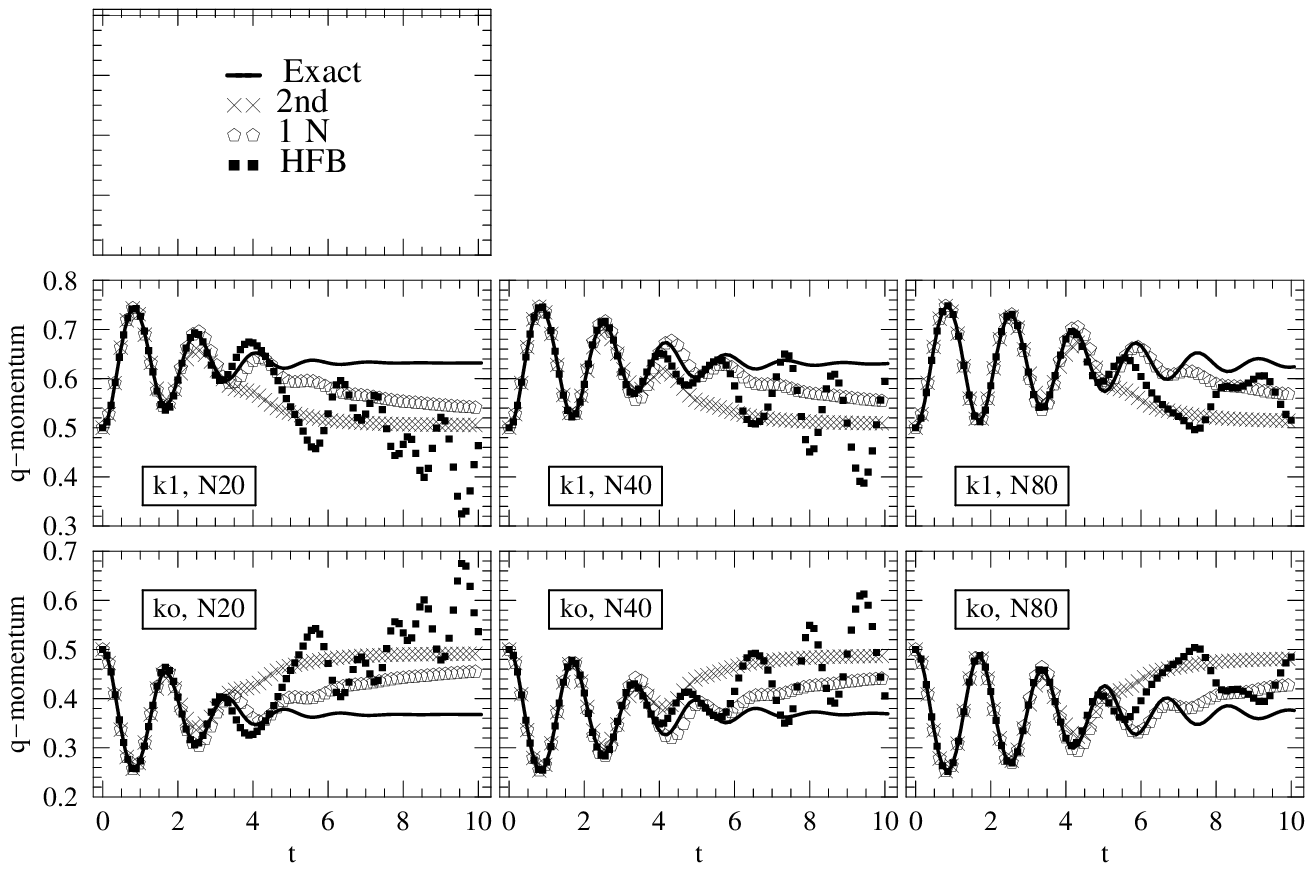}}
\end{center}
\caption{ Dynamical evolution of the quasi-momentum intensities. The
parameters used were $I=2,J=1/2$, $NU/J=4$ and $N=20,40$ and $80$. Time is
in units of $\hbar/J$. In the plots ko denotes the $k=0$ quasi-momentum
component and k1 the $k=\protect\pi/a$ one ($a$ the lattice spacing). The
plots are scaled to set the total quasi-momentum intensity to one for all
the different N cases. }
\label{Fig5c}
\end{figure}

In Figs. 3 to 7 we show our numerical results. We focus our attention on the
evolution of the condensate population per well, $|\phi _{i}(t)|^{2}$, the
total atomic population per well, $|\phi _{i}|^{2}+\rho _{ii}^{(F)}(t,t)-%
\frac{1}{2}$, the depletion per well or atoms out of the condensate, $\rho
_{ii}^{(F)}(t,t)-\frac{1}{2}$, and the total condensate population,$%
\sum_{i}|\phi _{i}(t)|^{2}$. The total population is also explicitly shown
in the figures to emphasize number conservation.

The quasi-momentum distribution of the atoms released from the lattice is
important because it is one of the most easily accessible quantities from an
experiment. The quasi-momentum distribution function $n_{k}$ is defined as
\begin{equation}
n_{k}(t)=\frac{1}{I}\sum_{i,j}e^{ik(i-j)}\left\langle \Phi
_{i}^{\dagger }(t)\Phi _{j}(t)\right\rangle,
\end{equation}

\noindent where the quasi-momentum $k$ can assume discrete values which are
integral multiples of $\frac{2\pi }{Ia},$ with $I$ the total number of
lattice sites and $a$ the lattice spacing.

The basic features of the plots can be summarized as follows:

\paragraph{In the very weak interacting regime}

(Fig. 3) the dynamics of the atomic population per well is almost
like Rabi oscillations. Notice that even though there are three
wells, periodic boundary conditions enforce \ equal evolution of
the initial empty ones. In this regime damping effects remains
very small for the time depicted in the plots. The numerical
simulations show a general agreement between the different
approaches with the exact solution. The effect of including higher
order terms in the equations of motion introduce small corrections
which improve the agreement with the exact dynamics. This shows up
in the plots of the condensate population and depletion, where the
small differences can be appreciated better. The second order
1/$\mathcal{N}$ expansion gives an improvement over the HFB and
the complete second order perturbative expansion almost matches
perfectly with the exact solution. In the duration depicted in the
plots of \ Fig. 3 the total condensate constitutes an important
fraction of the total population. Regarding the quasi-momentum
distribution we observe that similar to the \ spatial distribution
where \ the initial \ configuration and periodic boundary
conditions reduce the three well system to a double well one, they
enforce
equal evolution of the $\pm \frac{2\pi }{3}$ quasimomentum intensities. The $%
k=0$ and $\pm \frac{2\pi }{3}$ intensities oscillate with the same frequency
as the atomic population per well, both are also well described by the
approximations in consideration.

\paragraph{In the intermediate regime}

we can see the effect of the interactions in the dynamics. They modulate the
oscillations in the population per well and scatter the atoms out of the
condensate.

\begin{enumerate}
\item  In Figs. 4 we plot the numerical solution for the parameters $%
I=3,N=8,J=1$ and $U/J=1/3$. Different from the very weak
interacting regime only at the very early times all three
approximations are close to the exact solution. Even though any of
them are good after the first oscillation the HFB approximation is
the only one that fails to capture the exponential decrease of the
condensate population. This is expected, because even though this
approximation goes beyond mean field theory and takes into account
the most important scattering effects, it includes the effects of
collisions only indirectly through energy shifts, and breaks down
outside the collisionless regime where multiple-scattering effects
are important. In contrast, the exponential decay of the
condensate is present in the second order approximations. Non
local parts of the self-energy included in them encode scattering
effects responsible for damping. It is important to point out
that, even though we observe the collapse of the condensate
population, the total population is always conserved: As the
condensate population decreases, the number of atoms out of the
condensate increases.

\item  Comparing the two second order approaches we observe that the full
second order expansion gives a better description of the dynamics than the 1/%
$\mathcal{N}$ solution only in the regime when the perturbative solutions
are close to the exact dynamics. As soon as the third order terms start to
be important the large 1/$\mathcal{N}$ expansion gives a better qualitative
description. This behavior is going to be appreciated better in figs. 5 to 7
as the number of atoms is increased (see discussion bellow).

We observe as a general issue in this regime that, regardless of
the fact that the second order solutions capture well the damping
effects, as soon as the condensate population decreases to a small
percentage of the total population, they depart from the exact
dynamics: the second order approaches predict faster damping
rates. The overdamping is more severe in the dynamics of the
population per well than in the condensate dynamics. The failure
can be understood under the following lines of reasoning. At zero
temperature condensate atoms represent the most ''classical'' form
of a matter wave. When they decay, the role of quantum
correlations become more important. At this point the higher order
terms neglected in the second order approximations are the ones
that lead the dynamical behavior. Thus, to have a more accurate
description of the dynamics after the coherent matter field has
decayed one needs a better treatment of correlations.

Damping effects are also quite noticeable in the quantum evolution of the
quasi-momentum intensities. Similar to what happens to the spatial
observables, the HFB approximation fails completely to capture the damping
effects present in the evolution of the Fourier intensities whereas the
second order approaches overestimate them.

\item  In Figs. 5 to 7 we explore the effect of the total number of atoms on
the dynamics. In the plots we show the numerical solutions found for a
double well system with fix ratio $UN/J=4$ and three different values of N:
N=20,40 and 80. We present the results obtained for the evolution of the
atomic population per well in Fig. 5, the condensate population per well and
total condensate population in Fig. 6 and the quasi-momentum intensities in
Fig. 7. To make the comparisons easier we scaled the numerical results
obtained for the three different values of N by dividing them by the total
number of atoms. In this way for all the cases we start with an atomic
population of magnitude one in the initial populated well.

In the exact dynamics we see that as the number of atoms is
increased the damping effects occur at  slower rates. This feature
can be noticed in the quantum dynamics of all of the observables
depicted in the plots 5 to 7. The decrease of the damping rates as
the number of atoms is increased is not surprising because by
changing the number of atoms we affect the quantum coherence
properties of the system. As shown in Ref. \cite{Walls}, the
collapse time of the condensate population is approximately given by $%
t_{coll}\sim \frac{t_{rev}}{\sigma }$ where $\sigma $ is the variance of the
initial atomic distribution and ${t_{rev}}$ is the revival time which
depends on the detailed spectrum for the hamiltonian. In the kinetic energy
dominated regime ${t_{rev}}\sim hN/J$ (see Ref.\cite{Smerzi}), thus $%
t_{coll}\sim \frac{N}{\sigma }$. For our initial conditions the variance is
proportional to $\sigma =\sqrt{N}$ so $t_{coll}\sim \sqrt{N}$. Besides
damping rates, the qualitative behavior of the exact quantum dynamics is not
affected very much as the number of atoms is increased for a fixed $UN/J$.

The improvement of the 2PI approximations as N is increased, as a
result of the increase in the initial number of coherent atoms is
in fact observed in the plots. Even though the problem of
underdamping in the HFB approximation and the overdamping in the
second order approaches are not cured, as the number of atoms is
increased, we do observe a better matching with the full quantal
solution. The 1/$\mathcal{N}$ expansion shows the fastest
convergence. Perhaps this issue can be more easily observed in the
quasi-momentum distribution plots, Fig.7. The better agreement of the 1/$%
\mathcal{N}$ expansion relies on the fact that even though the number of
fields is only two in our calculations the 1/$\mathcal{N}$ expansion is an
expansion about a strong quasiclassical field configuration.
\end{enumerate}

\section{Conclusions}

In this work we have \ used the CTP functional formalism \ for 2PI Green's
functions to describe the nonequilibrium dynamics of a condensate loaded in
an optical lattice on every third lattice sites. We have carried out the
analysis up to second order in the interaction strength.This approximation
is introduced so as to make the numerical solution manageable, but it is
sufficient to account for dissipative effects due to multiparticle
scattering that are crucial \ even at early times. Our formulation is
capable of capturing the salient features of the system dynamics in the
regime under consideration, such as the decay of the condensate population
and the damping of the \ oscillations of the quasi-momentum and population
per well unaccounted for in the HFB approximation. However, at the point
where an important fraction of the condensate population has been scattered
out, the second order approximations used here predict an overdamped
dynamics. To improve on this a better treatment of higher correlations is
required. One might try to include the full next to leading order large $%
\mathcal{N}$ expansion without the truncation to second order as done in
Ref. \cite{Berges} but it is not obvious that this will lead to the required
improvement. Alternatively, one may try to adopt a stochastic approach, but
the challenge will be shifted to the derivation of a noise term (which is
likely to be both colored and multiplicative) which contains the effects of
these higher correlations and the solution of the stochastic equations. We
hope to address this aspect of the problem in a a future work.

Even though, as is clear in this paper, the second order 2PI approximations
fail to capture the fully correlated dynamics in the system, it has been
proved to work at intermediate times when correlations are not negligible
and standard mean field techniques fail poorly. Because of its success in
describing moderately correlated regimes, the second order approximations
could become a useful tool for describing experimental situations as the
collapsing or colliding condensates experiments where striking dynamical
behavior such as collisional loss of condensate atoms has been observed.

In summary, we have presented a new approach for the description of the
nonequilibrium dynamics of a Bose-Einstein condensate and fluctuations in a
closed quantum field system. The formalism allows one to go beyond the well
known HFB approximation and to incorporate the nonlinear and non-Markovian
aspects of the quantum dynamics as manifest in the dissipation and
fluctuations phenomena. The 2PI effective action formalism provides a useful
framework where the mean field and the correlation functions are treated on
the same footing self-consistently and that respects conservations of
particle number and energy. The CTP formalism ensures that the dynamical
equations of motion are also causal. In their current form scattering terms
non local in time are hard to estimate analytically and numerically
demanding. However, this systematic approach can be used as a quantitative
means to obtain solutions in different regimes and make comparisons with
kinetic theory results where a  Markovian approximation is assumed.

\paragraph{Acknowledgement}

A. M. Rey acknowledges correspondence with Dr. Berges on his recent large $%
\mathcal{N}$ numerical results and discussions with Ted Kirkpatrick. A.M.
Rey and C. Clark are supported in part by an Advanced Research and
Development Activity contract and by the U.S. National Science Foundation
under grant PHY-0100767. B. L. Hu and the visits of E. Calzetta are
supported in part by a NSF grant, a NIST grant and an ARDA contract. E.
Calzetta is supported by the University of Buenos Aires, CONICET, Fundacion
Antorchas and ANPCyT under project PICT 99 03-05229. A. Roura is supported
by an NSF under the grand PHY-9800967.

\renewcommand{\theequation}{A-\arabic{equation}}
\setcounter{equation}{0} 

\section*{Appendix 1: Mode Expansion of the HFB Equations}

To decouple the HFB equations we apply the well known Bogoliubov
transformation to the fluctuation field:

\begin{equation}
\varphi_{j}(t)=\sum_{q}u_{i}^{q}(t)\hat{\alpha}_{q}-v_{i}^{\ast q}(t)%
\hat{\alpha}_{q}^{\dagger },
\end{equation}
where $(\hat{\alpha}_{q},\hat{\alpha}_{q}^{\dagger })$ are time independent
creation and annihilation quasiparticle operators and all the time
dependence is absorbed in the amplitudes \ $\{u_{i}^{q}(t),v_{i}^{\ast
q}(t)\}.$ To ensure that the quasi-particle transformation is canonical, the
amplitudes $\{u_{i}^{q}(t),v_{i}^{\ast q}(t)\}$ \ have to fulfill the
following relations\footnote{%
The HFB approximation can be seen as a quadratic approximation of the
Hamiltonian in which third and quartic terms are reduced to linear and
quadratic forms by factorizing them in a self-consistent Gaussian
approximation. Any quadratic Hamiltonian can be diagonalized exactly and
this is achieved by transforming to a quasi-particle basis, $\{\hat{\alpha}%
_{q}\}$. The requirement of the transformation to be canonical means that it
preserves commutation relation and leads to bosonic quasi-particles.}:

\begin{eqnarray}
\sum_{i}u_{i}^{q}(t)u_{i}^{\ast k}(t)-v_{i}^{q}(t)v_{i}^{\ast k}(t)
&=&\delta _{qk},  \label{cons1} \\
\sum_{i}u_{i}^{q}(t)v_{i}^{k}(t)-v_{i}^{q}(t)u_{i}^{k}(t) &=&0.
\label{cons2}
\end{eqnarray}
In the zero temperature limit, where $\left\langle \hat{\alpha}_{q}^{\dagger
}\hat{\alpha}_{k}\right\rangle =0$, the statistical and spectral functions
take the form

\begin{eqnarray}
\rho _{ij}^{(F)}(t,t^{\prime }) &=&\frac{1}{2}\sum_{q}\left(
v_{i}^{q}(t)v_{j}^{\ast q}(t^{\prime })+u_{j}^{q}(t^{\prime })u_{i}^{\ast
q}(t)\right),  \label{rofhfb} \\
\rho _{ij}^{(\rho )}(t,t^{\prime }) &=&i\sum_{q}\left(
v_{i}^{q}(t)v_{j}^{\ast q}(t^{\prime })-u_{j}^{q}(t^{\prime })u_{i}^{\ast
q}(t)\right), \\
m_{ij}^{(F)}(t,t^{\prime }) &=&\frac{1}{2}\sum_{q}\left(
u_{i}^{q}(t)v_{j}^{\ast q}(t^{\prime })+u_{j}^{q}(t^{\prime })v_{i}^{\ast
q}(t)\right), \\
m_{ij}^{(\rho )}(t,t^{\prime }) &=&i\sum_{q}\left( u_{i}^{q}(t)v_{j}^{\ast
q}(t^{\prime })-u_{j}^{q}(t^{\prime })v_{i}^{\ast q}(t)\right).
\label{mrhfb}
\end{eqnarray}
Replacing Eqs. (\ref{rofhfb})-(\ref{mrhfb}) into Eqs.
(\ref{hfbc})-(\ref {hfbmm}) and using the constraints
(\ref{cons1})-(\ref{cons2}) we recover the standard time dependent
equations for the quasiparticle amplitudes \cite {Griffin}:

\begin{equation}
i\hbar \partial _{t}\phi _{i}(t)=-J(\phi _{i+1}(t)+\phi _{i-1}(t))+U(|\phi
_{i}(t)|^{2}+2\rho _{ii}^{(F)}(t,t))\phi _{i}(t)+Um_{ii}^{(F)}(t,t)\phi
_{i}^{\ast }(t),  \label{qchf}
\end{equation}

\begin{eqnarray}
i\hbar \frac{\partial }{\partial t}u_{i}^{q}(t)
&=&-J(u_{i+1}^{q}(t)+u_{i-1}^{q}(t))+2U(|\phi _{i}(t)|^{2}+\rho
_{ii}^{(F)}(t,t))u_{i}^{q}(t)-U(m_{ii}^{\ast (F)}(t,t)+\phi _{i}^{\ast
}(t)^{2})v_{i}^{q}(t), \\
-i\hbar \frac{\partial }{\partial t}v_{i}^{q}(t)
&=&-J(v_{i+1}^{q}(t)+v_{i-1}^{q}(t)))+2U(|\phi _{i}(t)|^{2}+\rho
_{ii}^{(F)}(t,t))v_{i}^{q}(t)-U(m_{ii}^{(F)}(t,t)+\phi
_{i}(t)^{2})u_{i}^{q}(t),  \label{hfbv}
\end{eqnarray}

\bigskip Equations \ (\ref{qchf}) -(\ref{hfbv}) correspond to a set of $%
I(2I+1)$ coupled ordinary differential equations, where $I$ is the total
number of lattice sites. They can be solved using standard time propagation
algorithms. Once the time dependent quasiparticle amplitudes is calculated\
we can derive the dynamics of \ physical observables constructed from them
as a function of time, such as the average number of particles in a well $%
n_i(t)$, etc.

\renewcommand{\theequation}{B-\arabic{equation}}
\setcounter{equation}{0} 

\section*{Appendix 2: Second order equations of Motion}

Hereunder we explicitly write the equations of motion of the $1/ \mathcal{N}$
and full second order approximations.

To simplify the notation let us introduce the functions

\begin{eqnarray}
\Omega _{ij}^{(F)}\left[ \mathbf{f},\mathbf{g}\right]  &=&\mathbf{f}%
_{ij}^{(F)}(t_{i},t_{j})\mathbf{g}_{ij}^{(F)}(t_{i},t_{j})-\frac{1}{4}\left(
\mathbf{f}_{ij}^{(\rho )}(t_{i},t_{j})\mathbf{g}_{ij}^{(\rho
)}(t_{i},t_{j})\right) , \\
\Omega _{ij}^{(\rho )}\left[ \mathbf{f},\mathbf{g}\right]  &=&\mathbf{f}%
_{ij}^{(F)}(t_{i},t_{j})\mathbf{g}_{ij}^{(\rho )}(t_{i},t_{j})+\mathbf{f}%
_{ij}^{(\rho )}(t,t^{\prime })\mathbf{g}_{ij}^{(F)}(t_{i},t_{j}).
\end{eqnarray}
Using the spectral and statistical functions and setting
$\mathcal{N}$ equal to 2 the equations of motion derived in Sec. 7
can be written as:

\begin{enumerate}
\item  Full second order expansion

If we take the complete contribution of the \emph{setting-sun} and \emph{%
basket-ball} diagrams the equations of motion take the form

\begin{eqnarray}
i\hbar \partial _{t_{i}}\phi _{i} &=&-J(\phi _{i+1}(t_{i})+\phi
_{i-1}(t_{i}))+U(|\phi _{i}|^{2}+2\rho _{ii}^{(F)})\phi
_{i}+Um_{ii}^{(F)}\phi _{i}^{\ast } \\
&&-2U^{2}\sum_{k}\int_{0}^{t_{i}}dt_{k}\left( \phi _{k}\Omega _{ik}^{(\rho
)} \left[ \rho ,\rho ^{\ast }\right] +\phi _{k}\Omega _{ik}^{(\rho )}\left[
m,m^{\ast }\right] +\phi _{k}^{\ast }\Omega _{ik}^{(\rho )}\left[ m,\rho %
\right] \right) \rho _{ki}^{(F)}  \notag \\
&&-2U^{2}\sum_{k}\int_{0}^{t_{i}}\left( \phi _{k}^{\ast }\Omega _{ik}^{(\rho
)}\left[ \rho ,\rho ^{\ast }\right] +\phi _{k}^{\ast }\Omega _{ik}^{(\rho )}%
\left[ m,m^{\ast }\right] +\phi _{k}^{{}}\Omega _{ik}^{(\rho )}\left[
m^{\ast },\rho ^{\ast }\right] \right) m_{ki}^{(F)}  \notag \\
&&+2U^{2}\sum_{k}\int_{0}^{t_{i}}dt_{k}\left( \phi _{k}\Omega _{ik}^{(F)}
\left[ \rho ,\rho ^{\ast }\right] +\phi _{k}\Omega _{ik}^{(F)}\left[
m,m^{\ast }\right] +\phi _{k}^{\ast }\Omega _{ik}^{(F)}\left[ m,\rho \right]
\right) \rho _{ki}^{(\rho )}  \notag \\
&&+2U^{2}\sum_{k}\int_{0}^{t_{i}}\left( \phi _{k}^{\ast }\Omega _{ik}^{(F)}
\left[ \rho ,\rho ^{\ast }\right] +\phi _{k}^{\ast }\Omega _{ik}^{(F)}\left[
m,m^{\ast }\right] +\phi _{k}^{{}}\Omega _{ik}^{(F)}\left[ m^{\ast },\rho
^{\ast }\right] \right) m_{ki}^{(\rho )},  \notag
\end{eqnarray}

\begin{eqnarray}
-i\hbar \partial _{t_{i}}\rho _{ij}^{(F)} &=&-J(\rho
_{i+1j}^{(F)}(t_{i},t_{j})+\rho _{i-1j}^{(F)}(t_{i},t_{j}))+2U(|\phi
_{i}|^{2}+\rho _{ii}^{(F)})\rho _{ij}^{(F)}+U(m_{ii}^{\ast (F)}+\phi
_{i}^{\ast 2})m_{ij}^{(F)} \\
&&-2U^{2}\sum_{k}\int_{0}^{t_{i}}dt_{k}\left( \phi _{i}\phi _{k}^{\ast
}\Omega _{ik}^{(\rho )}\left[ \rho ,\rho \right] +2\phi _{i}\phi
_{k}^{{}}\Omega _{ik}^{(\rho )}\left[ \rho ,m^{\ast }\right] +2\phi
_{i}^{\ast }\phi _{k}^{\ast }\Omega _{ik}^{(\rho )}\left[ m,\rho \right]
\right) \rho _{kj}^{(F)}  \notag \\
&&-2U^{2}\sum_{k}\int_{0}^{t_{i}}dt_{k}\left( \Omega _{ik}^{(\rho )}\left[
\rho ,\Delta \right] +2\phi _{i}^{\ast }\phi _{k}\left( \Omega _{ik}^{(\rho
)}\left[ \rho ,\rho ^{\ast }\right] +\Omega _{ik}^{(\rho )}\left[ m,m^{\ast }%
\right] \right) \right) \rho _{kj}^{(F)}  \notag \\
&&-2U^{2}\sum_{k}\int_{0}^{t_{i}}dt_{k}\left( 2\phi _{i}\phi _{k}^{\ast
}\Omega _{ik}^{(\rho )}\left[ m^{\ast },\rho \right] +\phi _{i}\phi
_{k}^{{}}\Omega _{ik}^{(\rho )}\left[ m^{\ast },m^{\ast }\right] +2\phi
_{i}^{\ast }\phi _{k}^{{}}\Omega _{ik}^{(\rho )}\left[ \rho ^{\ast },m^{\ast
}\right] \right) m_{kj}^{(F)}  \notag \\
&&-2U^{2}\sum_{k}\int_{0}^{t_{i}}dt_{k}\left( \Omega _{ik}^{(\rho )}\left[
m^{\ast },\Upsilon \right] +\phi _{i}^{\ast }\phi _{k}^{\ast }\left( 2\Omega
_{ik}^{(\rho )}\left[ \rho ,\rho ^{\ast }\right] +2\Omega _{ik}^{(\rho )}%
\left[ m,m^{\ast }\right] \right) \right) m_{kj}^{(F)}  \notag \\
&&+2U^{2}\sum_{k}\int_{0}^{t_{j}}dt_{k}\left( \phi _{i}\phi _{k}^{\ast
}\Omega _{ik}^{(F)}\left[ \rho ,\rho \right] +2\phi _{i}\phi _{k}^{{}}\Omega
_{ik}^{(F)}\left[ \rho ,m^{\ast }\right] +2\phi _{i}^{\ast }\phi _{k}^{\ast
}\Omega _{ik}^{(F)}\left[ m,\rho \right] \right) \rho _{kj}^{(\rho )}  \notag
\\
&&+2U^{2}\sum_{k}\int_{0}^{t_{j}}dt_{k}\left( \Omega _{ik}^{(F)}\left[ \rho
,\Delta \right] +2\phi _{i}^{\ast }\phi _{k}\left( \Omega _{ik}^{(F)}\left[
\rho ,\rho ^{\ast }\right] +\Omega _{ik}^{(F)}\left[ m,m^{\ast }\right]
\right) \right) \rho _{kj}^{(\rho )}  \notag \\
&&+2U^{2}\sum_{k}\int_{0}^{t_{j}}dt_{k}\left( 2\phi _{i}\phi _{k}^{\ast
}\Omega _{ik}^{(F)}\left[ m^{\ast },\rho \right] +\phi _{i}\phi
_{k}^{{}}\Omega _{ik}^{(F)}\left[ m^{\ast },m^{\ast }\right] +2\phi
_{i}^{\ast }\phi _{k}^{{}}\Omega _{ik}^{(F)}\left[ \rho ^{\ast },m^{\ast }%
\right] \right) m_{kj}^{(\rho )}  \notag \\
&&+2U^{2}\sum_{k}\int_{0}^{t_{j}}dt_{k}\left( \Omega _{ik}^{(F)}\left[
m^{\ast },\Upsilon \right] +2\phi _{i}^{\ast }\phi _{k}^{\ast }\left( \Omega
_{ik}^{(F)}\left[ \rho ,\rho ^{\ast }\right] +\Omega _{ik}^{(F)}\left[
m,m^{\ast }\right] \right) \right) m_{kj}^{(F)},  \notag
\end{eqnarray}

\begin{eqnarray}
-i\hbar \partial _{t_{i}}\rho _{ij}^{(\rho )} &=&-J(\rho _{i+1j}^{(\rho
)}(t_{i},t_{j})+\rho _{i-1j}^{(\rho )}(t_{i},t_{j}))+2U(|\phi _{i}|^{2}+\rho
_{ii}^{(F)})\rho _{ij}^{(\rho )}+U(m_{ii}^{\ast (F)}+\phi _{i}^{\ast
2})m_{ij}^{(\rho )} \\
&&-2U^{2}\sum_{k}\int_{t_{j}}^{t_{i}}dt_{k}\left( \phi _{i}\phi _{k}^{\ast
}\Omega _{ik}^{(\rho )}\left[ \rho ,\rho \right] +2\phi _{i}\phi
_{k}^{{}}\Omega _{ik}^{(\rho )}\left[ \rho ,m^{\ast }\right] +2\phi
_{i}^{\ast }\phi _{k}^{\ast }\Omega _{ik}^{(\rho )}\left[ m,\rho \right]
\right) \rho _{kj}^{(\rho )}  \notag \\
&&-2U^{2}\sum_{k}\int_{t_{j}}^{t_{i}}dt_{k}\left( \Omega _{ik}^{(\rho )}
\left[ \rho ,\Delta \right] +2\phi _{i}^{\ast }\phi _{k}\left( \Omega
_{ik}^{(\rho )}\left[ \rho ,\rho ^{\ast }\right] +\Omega _{ik}^{(\rho )}%
\left[ m,m^{\ast }\right] \right) \right) \rho _{kj}^{(\rho )}  \notag \\
&&-2U^{2}\sum_{k}\int_{t_{j}}^{t_{i}}dt_{k}\left( 2\phi _{i}\phi _{k}^{\ast
}\Omega _{ik}^{(\rho )}\left[ m^{\ast },\rho \right] +\phi _{i}\phi
_{k}^{{}}\Omega _{ik}^{(\rho )}\left[ m^{\ast },m^{\ast }\right] +2\phi
_{i}^{\ast }\phi _{k}^{{}}\Omega _{ik}^{(\rho )}\left[ \rho ^{\ast },m^{\ast
}\right] \right) m_{kj}^{(\rho )}  \notag \\
&&-2U^{2}\sum_{k}\int_{t_{j}}^{t_{i}}dt_{k}\left( \Omega _{ik}^{(\rho )}
\left[ m^{\ast },\Upsilon \right] +2\phi _{i}^{\ast }\phi _{k}^{\ast }\left(
\Omega _{ik}^{(\rho )}\left[ \rho ,\rho ^{\ast }\right] +\Omega _{ik}^{(\rho
)}\left[ m,m^{\ast }\right] \right) \right) m_{kj}^{(\rho )},  \notag
\end{eqnarray}

\begin{eqnarray}
i\hbar \partial _{t_{i}}m_{ij}^{(F)}
&=&-J(m_{i+1j}^{(F)}(t_{i},t_{j})+m_{i-1j}^{(F)}(t_{i},t_{j}))+2U(|\phi
_{i}|^{2}+\rho _{ii}^{(F)})m_{ij}^{(F)}+U(m_{ii}^{(F)}+\phi _{i}^{2})\rho
_{ij}^{(F)} \\
&&-2U^{2}\sum_{k}\int_{0}^{t_{i}}dt_{k}\left( 2\phi _{i}^{\ast }\phi
_{k}^{{}}\Omega _{ik}^{(\rho )}\left[ m,\rho ^{\ast }\right] +\phi
_{i}^{\ast }\phi _{k}^{\ast }\Omega _{ik}^{(\rho )}\left[ m,m\right] +2\phi
_{i}^{{}}\phi _{k}^{\ast }\Omega _{ik}^{(\rho )}\left[ \rho ,m\right]
\right) \rho _{kj}^{(F)}  \notag \\
&&-2U^{2}\sum_{k}\int_{0}^{t_{i}}dt_{k}\left( \Omega _{ik}^{(\rho )}\left[
m,\Upsilon \right] +2\phi _{i}^{{}}\phi _{k}\left( \Omega _{ik}^{(\rho )}%
\left[ \rho ,\rho ^{\ast }\right] +\Omega _{ik}^{(\rho )}\left[ m,m^{\ast }%
\right] \right) \right) \rho _{kj}^{(F)}  \notag \\
&&-2U^{2}\sum_{k}\int_{0}^{t_{i}}dt_{k}\left( \phi _{i}^{\ast }\phi
_{k}^{{}}\Omega _{ik}^{(\rho )}\left[ \rho ^{\ast },\rho ^{\ast }\right]
+2\phi _{i}^{\ast }\phi _{k}^{\ast }\Omega _{ik}^{(\rho )}\left[ \rho ^{\ast
},m\right] +2\phi _{i}^{{}}\phi _{k}^{{}}\Omega _{ik}^{(\rho )}\left[
m^{\ast },\rho ^{\ast }\right] \right) m_{kj}^{(F)}  \notag \\
&&-2U^{2}\sum_{k}\int_{0}^{t_{i}}dt_{k}\left( \Omega _{ik}^{(\rho )}\left[
\rho ^{\ast },\Delta \right] +2\phi _{i}^{{}}\phi _{k}^{\ast }\left( \Omega
_{ik}^{(\rho )}\left[ \rho ,\rho ^{\ast }\right] +\Omega _{ik}^{(\rho )}%
\left[ m,m^{\ast }\right] \right) \right) m_{kj}^{(F)}  \notag \\
&&+2U^{2}\sum_{k}\int_{0}^{t_{j}}dt_{k}\left( 2\phi _{i}^{\ast }\phi
_{k}^{{}}\Omega _{ik}^{(F)}\left[ m,\rho ^{\ast }\right] +\phi _{i}^{\ast
}\phi _{k}^{\ast }\Omega _{ik}^{(F)}\left[ m,m\right] +2\phi _{i}^{{}}\phi
_{k}^{\ast }\Omega _{ik}^{(F)}\left[ \rho ,m\right] \right) \rho
_{kj}^{(\rho )}  \notag \\
&&+2U^{2}\sum_{k}\int_{0}^{t_{j}}dt_{k}\left( \Omega _{ik}^{(F)}\left[
m,\Upsilon \right] +2\phi _{i}^{{}}\phi _{k}\left( \Omega _{ik}^{(F)}\left[
\rho ,\rho ^{\ast }\right] +\Omega _{ik}^{(F)}\left[ m,m^{\ast }\right]
\right) \right) \rho _{kj}^{(\rho )}  \notag \\
&&+2U^{2}\sum_{k}\int_{0}^{t_{j}}dt_{k}\left( \phi _{i}^{\ast }\phi
_{k}^{{}}\Omega _{ik}^{(F)}\left[ \rho ^{\ast },\rho ^{\ast }\right] +2\phi
_{i}^{\ast }\phi _{k}^{\ast }\Omega _{ik}^{(F)}\left[ \rho ^{\ast },m\right]
+2\phi _{i}^{{}}\phi _{k}^{{}}\Omega _{ik}^{(F)}\left[ m^{\ast },\rho ^{\ast
}\right] \right) m_{kj}^{(\rho )}  \notag \\
&&+2U^{2}\sum_{k}\int_{0}^{t_{j}}dt_{k}\left( \Omega _{ik}^{(F)}\left[ \rho
^{\ast },\Delta \right] +2\phi _{i}^{{}}\phi _{k}^{\ast }\left( \Omega
_{ik}^{(F)}\left[ \rho ,\rho ^{\ast }\right] +\Omega _{ik}^{(F)}\left[
m,m^{\ast }\right] \right) \right) m_{kj}^{(\rho )},  \notag
\end{eqnarray}

\begin{eqnarray}
i\hbar \partial _{t_{i}}m_{ij}^{(\rho )} &=&-J(m_{i+1j}^{(\rho
)}(t_{i},t_{j})+m_{i-1j}^{(\rho )}(t_{i},t_{j}))+2U(|\phi _{i}|^{2}+\rho
_{ii}^{(F)})m_{ij}^{(\rho )}+U(m_{ii}^{(F)}+\phi _{i}^{2})\rho _{ij}^{(\rho
)} \\
&&+2U^{2}\sum_{k}\int_{t_{j}}^{t_{i}}dt_{k}\left( 2\phi _{i}^{\ast }\phi
_{k}^{{}}\Omega _{ik}^{(\rho )}\left[ m,\rho ^{\ast }\right] +\phi
_{i}^{\ast }\phi _{k}^{\ast }\Omega _{ik}^{(\rho )}\left[ m,m\right] +2\phi
_{i}^{{}}\phi _{k}^{\ast }\Omega _{ik}^{(\rho )}\left[ \rho ,m\right]
\right) \rho _{kj}^{(\rho )}  \notag \\
&&+2U^{2}\sum_{k}\int_{t_{j}}^{t_{i}}dt_{k}\left( \Omega _{ik}^{(\rho )}
\left[ m,\Upsilon \right] +\phi _{i}^{{}}\phi _{k}\left( 2\Omega
_{ik}^{(\rho )}\left[ \rho ,\rho ^{\ast }\right] +2\Omega _{ik}^{(\rho )}%
\left[ m,m^{\ast }\right] \right) \right) \rho _{kj}^{(\rho )}  \notag \\
&&+2U^{2}\sum_{k}\int_{t_{j}}^{t_{i}}dt_{k}\left( \phi _{i}^{\ast }\phi
_{k}^{{}}\Omega _{ik}^{(\rho )}\left[ \rho ^{\ast },\rho ^{\ast }\right]
+2\phi _{i}^{\ast }\phi _{k}^{\ast }\Omega _{ik}^{(\rho )}\left[ \rho ^{\ast
},m\right] +2\phi _{i}^{{}}\phi _{k}^{{}}\Omega _{ik}^{(\rho )}\left[
m^{\ast },\rho ^{\ast }\right] \right) m_{kj}^{(\rho )}  \notag \\
&&+2U^{2}\sum_{k}\int_{t_{j}}^{t_{i}}dt_{k}\left( \Omega _{ik}^{(\rho )}
\left[ \rho ^{\ast },\Delta \right] +2\phi _{i}^{{}}\phi _{k}^{\ast }\left(
\Omega _{ik}^{(\rho )}\left[ \rho ,\rho ^{\ast }\right] +\Omega _{ik}^{(\rho
)}\left[ m,m^{\ast }\right] \right) \right) m_{kj}^{(\rho )},  \notag
\end{eqnarray}
with
\begin{eqnarray}
\Delta _{ij}^{(F,\rho )} &=&\Omega _{ij}^{(F,\rho )}\left[ \rho ,\rho ^{\ast
}\right] +2\Omega _{ij}^{(F,\rho )}\left[ m,m^{\ast }\right] , \\
\Upsilon _{ij}^{(F,\rho )} &=&2\Omega _{ij}^{(F,\rho )}\left[ \rho ,\rho
^{\ast }\right] +\Omega _{ij}^{(F,\rho )}\left[ m,m^{\ast }\right] .
\end{eqnarray}

\item  Second order $1/\mathcal{N}$ expansion:
\begin{eqnarray}
i\hbar \partial _{t_{i}}\phi _{i} &=&-J(\phi _{i+1}(t_{i})+\phi
_{i-1}(t_{i}))+U(|\phi _{i}|^{2}+2\rho _{ii}^{(F)})\phi
_{i}+Um_{ii}^{(F)}\phi _{i}^{\ast } \\
&&-U^{2}\sum_{k}\int_{0}^{t_{i}}dt_{k}\Pi _{ik}^{(\rho )}\left( \phi
_{k}\rho _{ki}^{(F)}+\phi _{k}^{\ast }m_{ki}^{(F)}\right)
+U^{2}\sum_{k}\int_{0}^{t_{i}}dt_{k}\Pi _{ik}^{(F)}\left( \phi _{k}\rho
_{ki}^{(\rho )}+\phi _{k}^{\ast }m_{ki}^{(\rho )}\right) ,  \notag \\
-i\hbar \partial _{t_{i}}\rho _{ij}^{(F)} &=&-J(\rho
_{i+1j}^{(F)}(t_{i},t_{j})+\rho _{i-1j}^{(F)}(t_{i},t_{j}))+2U(|\phi
_{i}|^{2}+\rho _{ii}^{(F)})\rho _{ij}^{(F)}+U(m_{ii}^{\ast (F)}+\phi
_{i}^{\ast 2})m_{ij}^{(F)} \\
&&-U^{2}\sum_{k}\int_{0}^{t_{i}}dt_{k}\left( \phi _{i}\phi _{k}^{\ast
}\Omega _{ik}^{(\rho )}\left[ \rho ,\rho \right] +\phi _{i}\phi
_{k}^{{}}\Omega _{ik}^{(\rho )}\left[ \rho ,m^{\ast }\right] +\phi
_{i}^{\ast }\phi _{k}^{\ast }\Omega _{ik}^{(\rho )}\left[ m,\rho \right]
\right) \rho _{kj}^{(F)}  \notag \\
&&-U^{2}\sum_{k}\int_{0}^{t_{i}}dt_{k}\left( \Omega _{ik}^{(\rho )}\left[
\rho ,\Pi \right] +\phi _{i}^{\ast }\phi _{k}\left( 2\Omega _{ik}^{(\rho )}%
\left[ \rho ,\rho ^{\ast }\right] +\Omega _{ik}^{(\rho )}\left[ m,m^{\ast }%
\right] \right) \right) \rho _{kj}^{(F)}  \notag \\
&&-U^{2}\sum_{k}\int_{0}^{t_{i}}dt_{k}\left( \phi _{i}\phi _{k}^{\ast
}\Omega _{ik}^{(\rho )}\left[ m^{\ast },\rho \right] +\phi _{i}\phi
_{k}^{{}}\Omega _{ik}^{(\rho )}\left[ m^{\ast },m^{\ast }\right] +\phi
_{i}^{\ast }\phi _{k}^{{}}\Omega _{ik}^{(\rho )}\left[ \rho ^{\ast },m^{\ast
}\right] \right) m_{kj}^{(F)}  \notag \\
&&-U^{2}\sum_{k}\int_{0}^{t_{i}}dt_{k}\left( \Omega _{ik}^{(\rho )}\left[
m^{\ast },\Pi \right] +\phi _{i}^{\ast }\phi _{k}^{\ast }\left( \Omega
_{ik}^{(\rho )}\left[ \rho ,\rho ^{\ast }\right] +2\Omega _{ik}^{(\rho )}%
\left[ m,m^{\ast }\right] \right) \right) m_{kj}^{(F)}  \notag \\
&&+U^{2}\sum_{k}\int_{0}^{t_{j}}dt_{k}\left( \phi _{i}\phi _{k}^{\ast
}\Omega _{ik}^{(F)}\left[ \rho ,\rho \right] +\phi _{i}\phi _{k}^{{}}\Omega
_{ik}^{(F)}\left[ \rho ,m^{\ast }\right] +\phi _{i}^{\ast }\phi _{k}^{\ast
}\Omega _{ik}^{(F)}\left[ m,\rho \right] \right) \rho _{kj}^{(\rho )}  \notag
\\
&&+U^{2}\sum_{k}\int_{0}^{t_{j}}dt_{k}\left( \Omega _{ik}^{(F)}\left[ \rho
,\Pi \right] +\phi _{i}^{\ast }\phi _{k}\left( 2\Omega _{ik}^{(F)}\left[
\rho ,\rho ^{\ast }\right] +\Omega _{ik}^{(F)}\left[ m,m^{\ast }\right]
\right) \right) \rho _{kj}^{(\rho )}  \notag \\
&&+U^{2}\sum_{k}\int_{0}^{t_{j}}dt_{k}\left( \phi _{i}\phi _{k}^{\ast
}\Omega _{ik}^{(F)}\left[ m^{\ast },\rho \right] +\phi _{i}\phi
_{k}^{{}}\Omega _{ik}^{(F)}\left[ m^{\ast },m^{\ast }\right] +\phi
_{i}^{\ast }\phi _{k}^{{}}\Omega _{ik}^{(F)}\left[ \rho ^{\ast },m^{\ast }%
\right] \right) m_{kj}^{(\rho )}  \notag \\
&&+U^{2}\sum_{k}\int_{0}^{t_{j}}dt_{k}\left( \Omega _{ik}^{(F)}\left[
m^{\ast },\Pi \right] +\phi _{i}^{\ast }\phi _{k}^{\ast }\left( \Omega
_{ik}^{(F)}\left[ \rho ,\rho ^{\ast }\right] +2\Omega _{ik}^{(F)}\left[
m,m^{\ast }\right] \right) \right) m_{kj}^{(F)},  \notag
\end{eqnarray}

\begin{eqnarray}
-i\hbar \partial _{t_{i}}\rho _{ij}^{(\rho )} &=&-J(\rho _{i+1j}^{(\rho
)}(t_{i},t_{j})+\rho _{i-1j}^{(\rho )}(t_{i},t_{j}))+2U(|\phi _{i}|^{2}+\rho
_{ii}^{(F)})\rho _{ij}^{(\rho )}+U(m_{ii}^{\ast (F)}+\phi _{i}^{\ast
2})m_{ij}^{(\rho )} \\
&&-U^{2}\sum_{k}\int_{t_{j}}^{t_{i}}dt_{k}\left( \phi _{i}\phi _{k}^{\ast
}\Omega _{ik}^{(\rho )}\left[ \rho ,\rho \right] +\phi _{i}\phi
_{k}^{{}}\Omega _{ik}^{(\rho )}\left[ \rho ,m^{\ast }\right] +\phi
_{i}^{\ast }\phi _{k}^{\ast }\Omega _{ik}^{(\rho )}\left[ m,\rho \right]
\right) \rho _{kj}^{(\rho )}  \notag \\
&&-U^{2}\sum_{k}\int_{t_{j}}^{t_{i}}dt_{k}\left( \Omega _{ik}^{(\rho )}\left[
\rho ,\Pi \right] +\phi _{i}^{\ast }\phi _{k}\left( 2\Omega _{ik}^{(\rho )}%
\left[ \rho ,\rho ^{\ast }\right] +\Omega _{ik}^{(\rho )}\left[ m,m^{\ast }%
\right] \right) \right) \rho _{kj}^{(\rho )}  \notag \\
&&-U^{2}\sum_{k}\int_{t_{j}}^{t_{i}}dt_{k}\left( \phi _{i}\phi _{k}^{\ast
}\Omega _{ik}^{(\rho )}\left[ m^{\ast },\rho \right] +\phi _{i}\phi
_{k}^{{}}\Omega _{ik}^{(\rho )}\left[ m^{\ast },m^{\ast }\right] +\phi
_{i}^{\ast }\phi _{k}^{{}}\Omega _{ik}^{(\rho )}\left[ \rho ^{\ast },m^{\ast
}\right] \right) m_{kj}^{(\rho )}  \notag \\
&&-U^{2}\sum_{k}\int_{t_{j}}^{t_{i}}dt_{k}\left( \Omega _{ik}^{(\rho )}\left[
m^{\ast },\Pi \right] +\phi _{i}^{\ast }\phi _{k}^{\ast }\left( \Omega
_{ik}^{(\rho )}\left[ \rho ,\rho ^{\ast }\right] +2\Omega _{ik}^{(\rho )}%
\left[ m,m^{\ast }\right] \right) \right) m_{kj}^{(\rho )},  \notag
\end{eqnarray}

\begin{eqnarray}
i\hbar \partial _{t_{i}}m_{ij}^{(F)}
&=&-J(m_{i+1j}^{(F)}(t_{i},t_{j})+m_{i-1j}^{(F)}(t_{i},t_{j}))+2U(|\phi
_{i}|^{2}+\rho _{ii}^{(F)})m_{ij}^{(F)}+U(m_{ii}^{(F)}+\phi _{i}^{2})\rho
_{ij}^{(F)} \\
&&-U^{2}\sum_{k}\int_{0}^{t_{i}}dt_{k}\left( \phi _{i}^{\ast }\phi
_{k}^{{}}\Omega _{ik}^{(\rho )}\left[ m,\rho ^{\ast }\right] +\phi
_{i}^{\ast }\phi _{k}^{\ast }\Omega _{ik}^{(\rho )}\left[ m,m\right] +\phi
_{i}^{{}}\phi _{k}^{\ast }\Omega _{ik}^{(\rho )}\left[ \rho ,m\right]
\right) \rho _{kj}^{(F)}  \notag \\
&&-U^{2}\sum_{k}\int_{0}^{t_{i}}dt_{k}\left( \Omega _{ik}^{(\rho )}\left[
m,\Pi \right] +\phi _{i}^{{}}\phi _{k}\left( \Omega _{ik}^{(\rho )}\left[
\rho ,\rho ^{\ast }\right] +2\Omega _{ik}^{(\rho )}\left[ m,m^{\ast }\right]
\right) \right) \rho _{kj}^{(F)}  \notag \\
&&-U^{2}\sum_{k}\int_{0}^{t_{i}}dt_{k}\left( \phi _{i}^{\ast }\phi
_{k}^{{}}\Omega _{ik}^{(\rho )}\left[ \rho ^{\ast },\rho ^{\ast }\right]
+\phi _{i}^{\ast }\phi _{k}^{\ast }\Omega _{ik}^{(\rho )}\left[ \rho ^{\ast
},m\right] +\phi _{i}^{{}}\phi _{k}^{{}}\Omega _{ik}^{(\rho )}\left[ m^{\ast
},\rho ^{\ast }\right] \right) m_{kj}^{(F)}  \notag \\
&&-U^{2}\sum_{k}\int_{0}^{t_{i}}dt_{k}\left( \Omega _{ik}^{(\rho )}\left[
\rho ^{\ast },\Pi \right] +\phi _{i}^{{}}\phi _{k}^{\ast }\left( 2\Omega
_{ik}^{(\rho )}\left[ \rho ,\rho ^{\ast }\right] +\Omega _{ik}^{(\rho )}%
\left[ m,m^{\ast }\right] \right) \right) m_{kj}^{(F)}  \notag \\
&&+U^{2}\sum_{k}\int_{0}^{t_{j}}dt_{k}\left( \phi _{i}^{\ast }\phi
_{k}^{{}}\Omega _{ik}^{(F)}\left[ m^{{}},\rho ^{\ast }\right] +\phi
_{i}^{\ast }\phi _{k}^{\ast }\Omega _{ik}^{(F)}\left[ m,m\right] +\phi
_{i}^{{}}\phi _{k}^{\ast }\Omega _{ik}^{(F)}\left[ \rho ,m\right] \right)
\rho _{kj}^{(\rho )}  \notag \\
&&+U^{2}\sum_{k}\int_{0}^{t_{j}}dt_{k}\left( \Omega _{ik}^{(F)}\left[ m,\Pi %
\right] +\phi _{i}^{{}}\phi _{k}\left( 2\Omega _{ik}^{(F)}\left[ \rho ,\rho
^{\ast }\right] +\Omega _{ik}^{(F)}\left[ m,m^{\ast }\right] \right) \right)
\rho _{kj}^{(\rho )}  \notag \\
&&+U^{2}\sum_{k}\int_{0}^{t_{j}}dt_{k}\left( \phi _{i}^{\ast }\phi
_{k}^{{}}\Omega _{ik}^{(F)}\left[ \rho ^{\ast },\rho ^{\ast }\right] +\phi
_{i}^{\ast }\phi _{k}^{\ast }\Omega _{ik}^{(F)}\left[ \rho ^{\ast },m\right]
+\phi _{i}^{{}}\phi _{k}^{{}}\Omega _{ik}^{(F)}\left[ m^{\ast },\rho ^{\ast }%
\right] \right) m_{kj}^{(\rho )}  \notag \\
&&+U^{2}\sum_{k}\int_{0}^{t_{j}}dt_{k}\left( \Omega _{ik}^{(F)}\left[ \rho
^{\ast },\Pi \right] +\phi _{i}^{{}}\phi _{k}^{\ast }\left( 2\Omega
_{ik}^{(F)}\left[ \rho ,\rho ^{\ast }\right] +\Omega _{ik}^{(F)}\left[
m,m^{\ast }\right] \right) \right) m_{kj}^{(\rho )},  \notag
\end{eqnarray}

\begin{eqnarray}
i\hbar \partial _{t_{i}}m_{ij}^{(\rho )} &=&-J(m_{i+1j}^{(\rho
)}(t_{i},t_{j})+m_{i-1j}^{(\rho )}(t_{i},t_{j}))+2U(|\phi _{i}|^{2}+\rho
_{ii}^{(F)})m_{ij}^{(\rho )}+U(m_{ii}^{(F)}+\phi _{i}^{2})\rho _{ij}^{(\rho
)} \\
&&+U^{2}\sum_{k}\int_{t_{j}}^{t_{i}}dt_{k}\left( \phi _{i}^{\ast }\phi
_{k}^{{}}\Omega _{ik}^{(\rho )}\left[ m,\rho ^{\ast }\right] +\phi
_{i}^{\ast }\phi _{k}^{\ast }\Omega _{ik}^{(\rho )}\left[ m,m\right] +\phi
_{i}^{{}}\phi _{k}^{\ast }\Omega _{ik}^{(\rho )}\left[ \rho ,m\right]
\right) \rho _{kj}^{(\rho )}  \notag \\
&&+U^{2}\sum_{k}\int_{t_{j}}^{t_{i}}dt_{k}\left( \Omega _{ik}^{(\rho )}\left[
m,\Pi \right] +\phi _{i}^{{}}\phi _{k}\left( \Omega _{ik}^{(\rho )}\left[
\rho ,\rho ^{\ast }\right] +2\Omega _{ik}^{(\rho )}\left[ m,m^{\ast }\right]
\right) \right) \rho _{kj}^{(\rho )}  \notag \\
&&+U^{2}\sum_{k}\int_{t_{j}}^{t_{i}}dt_{k}\left( \phi _{i}^{\ast }\phi
_{k}^{{}}\Omega _{ik}^{(\rho )}\left[ \rho ^{\ast },\rho ^{\ast }\right]
+\phi _{i}^{\ast }\phi _{k}^{\ast }\Omega _{ik}^{(\rho )}\left[ \rho ^{\ast
},m\right] +\phi _{i}^{{}}\phi _{k}^{{}}\Omega _{ik}^{(\rho )}\left[ m^{\ast
},\rho ^{\ast }\right] \right) m_{kj}^{(\rho )}  \notag \\
&&+U^{2}\sum_{k}\int_{t_{j}}^{t_{i}}dt_{k}\left( \Omega _{ik}^{(\rho )}\left[
\rho ^{\ast },\Pi \right] +\phi _{i}^{{}}\phi _{k}^{\ast }\left( 2\Omega
_{ik}^{(\rho )}\left[ \rho ,\rho ^{\ast }\right] +\Omega _{ik}^{(\rho )}%
\left[ m,m^{\ast }\right] \right) \right) m_{kj}^{(\rho )},  \notag
\end{eqnarray}
with

\begin{equation}
\Pi _{ij}^{(F,\rho )}=\Omega _{ij}^{(F,\rho )}\left[ \rho ,\rho ^{\ast }%
\right] +\Omega _{ij}^{(F,\rho )}\left[ m,m^{\ast }\right] .
\end{equation}
In the above equations we have simplified the notation replacing $\phi
_{k}(t_{k})$ by $\phi _{k}$ and $m_{kj}(t_{k},t_{j})$ by $m_{kj}.$
\end{enumerate}

\end{document}